\documentclass{pasa}%

\title[]{Accuracy of Shack-Hartmann wavefront sensor using a coherent wound fibre image bundle}
\author[Author1 et al.]{Jessica R. Zheng, Michael Goodwin, Jon Lawrence \\
\affil{Australian Astronomical Observatory, PO Box 915, North Ryde, NSW 1670, Australia}%
}%
\jid{PASA}
\doi{10.1017/pas.\the\year.xxx}
\jyear{\the\year}

\usepackage[authoryear]{natbib}
\bibpunct{(}{)}{;}{a}{}{,}
\usepackage{aas_macros}
\usepackage{hyperref} 
\hypersetup{colorlinks,citecolor=blue,linkcolor=blue,urlcolor=blue}
\hypersetup{draft}

\begin{document}

\begin{abstract}
Shack-Hartmann wavefront sensors using wound fibre bundles are desired for multi-object adaptive optical systems to provide large multiplex positioned by Starbugs. \textcolor{black}{The use of the large-sized wound fibre bundle provides the flexibility to use more sub-apertures wavefront sensor for ELTs.} These compact wavefront sensors take advantage of large focal surfaces such as the Giant Magellan Telescope. The focus of this paper is to study the wound fibre image bundle structure defects effect on the centroid measurement accuracy of a Shack-Hartmann wavefront sensor. We use the first moment centroid method to estimate the centroid of a focused Gaussian beam sampled by a simulated bundle. Spot estimation accuracy with wound fibre image bundle and its structure impact on wavefront measurement accuracy statistics are addressed. Our results show that when the measurement signal to noise ratio is high, the centroid measurement accuracy is dominated by the wound fibre image bundle structure, e.g. tile angle and gap spacing. For the measurement with low signal to noise ratio, its accuracy is influenced by the read noise of the detector instead of the wound fibre image bundle structure defects. We demonstrate this both with simulation and experimentally. We provide a statistical model of the centroid and wavefront error of a wound fibre image bundle found through experiment. 
\end{abstract}

\maketitle

\section{INTRODUCTION}
\label{sec:intro}
Shack$-$Hartmann wavefront sensor (SHWFS) is a widely applied technique for measuring wavefront aberrations for adaptive optics (AO) (\cite{Hardy1998}). A fibre image bundle based SHWFS could enable a more robust and flexible wavefront measurement for extremely large telescope that are currently being built or proposed. In particular, Multi-object adaptive optics (MOAO) systems can maximise the scientific output of those large aperture wide-field telescopes, like the Giant Magellan Telescope (GMT). For MOAO to be realized on telescopes,  many wavefront sensors are needed to be deployed across the focal plane, localizing at each science target. We proposed two types of wavefront sensor by using fibre image bundles (\cite{Goodwin2014, Zheng2014}) where the image formed on the wavefront sensor can be relayed by a fibre image bundle to a more convenient location. The concept of an image bundle based SHWFS for Starbugs has been detailed by \cite{Goodwin2014} with an early lab prototype built for the GMT \cite{Goodwin2015}. The advantage of these miniature wavefront sensors is that each device can be fit in one $'$Starbug$'$ fibre positioning device currently under development at Australian Astronomical Observatory (\cite{Gilbert2012}, \cite{Staszak2016}). Multiple miniature wavefront sensors can be multiplexed to a common low noise camera and deployed in multiple locations according to the available natural or laser guide stars on the telescope focal plane and position reconfigured to meet different observations. It can potentially be used for ground-layer AO in which a single deformable mirror for ground atmospheric disturbance correction can be controlled by averaging the signals from multiple wavefront sensors pointing in widely separated directions(\cite{Ammons2010, Ono2016}).  Another application for image bundle based SHWFS is active optics, where a large number of wavefront sensors can help with the mirror alignment under slow changing conditions.

In our previous two publications, the fibre image bundle discussed is the polymer coherent fibre image bundle. Its performance(\cite{Richards2017}) is \textcolor{black}{comparable to the conventional wavefront sensor} however one major shortcoming is its rigidity and transmission loss, particularly in the near infra-red. For large telescope wavefront sensing, the relaying image fibre bundle needs to be more than a few millimetres in diameter to allow for size of the needed microlens array (e.g. 50x50 apertures \textcolor{black}{to accommodate more sub-apertures within the telescope pupil to be able to measure the wavefront more precisely}) .  A polymer fibre image bundle of this size  won$'$t  be able to move around freely due to bundle rigidity.  Other than the polymer fibre image bundle, there are some other flexible image fibre bundles currently available on the market(\cite{Schott2007}) such as the Schott leached fibre image bundle and the wound fibre image bundle. The leached fibre image bundles have a maximum diameter of about 1.5 mm and therefore are not suitable for large aperture wavefront sensing. However, The wound fibre image bundle can easily be scaled up to tens of millimetres and hence the best candidate for a Starbug WFS. 

\begin{figure*}
	\begin{center}
		\includegraphics[width= 1.8\columnwidth]{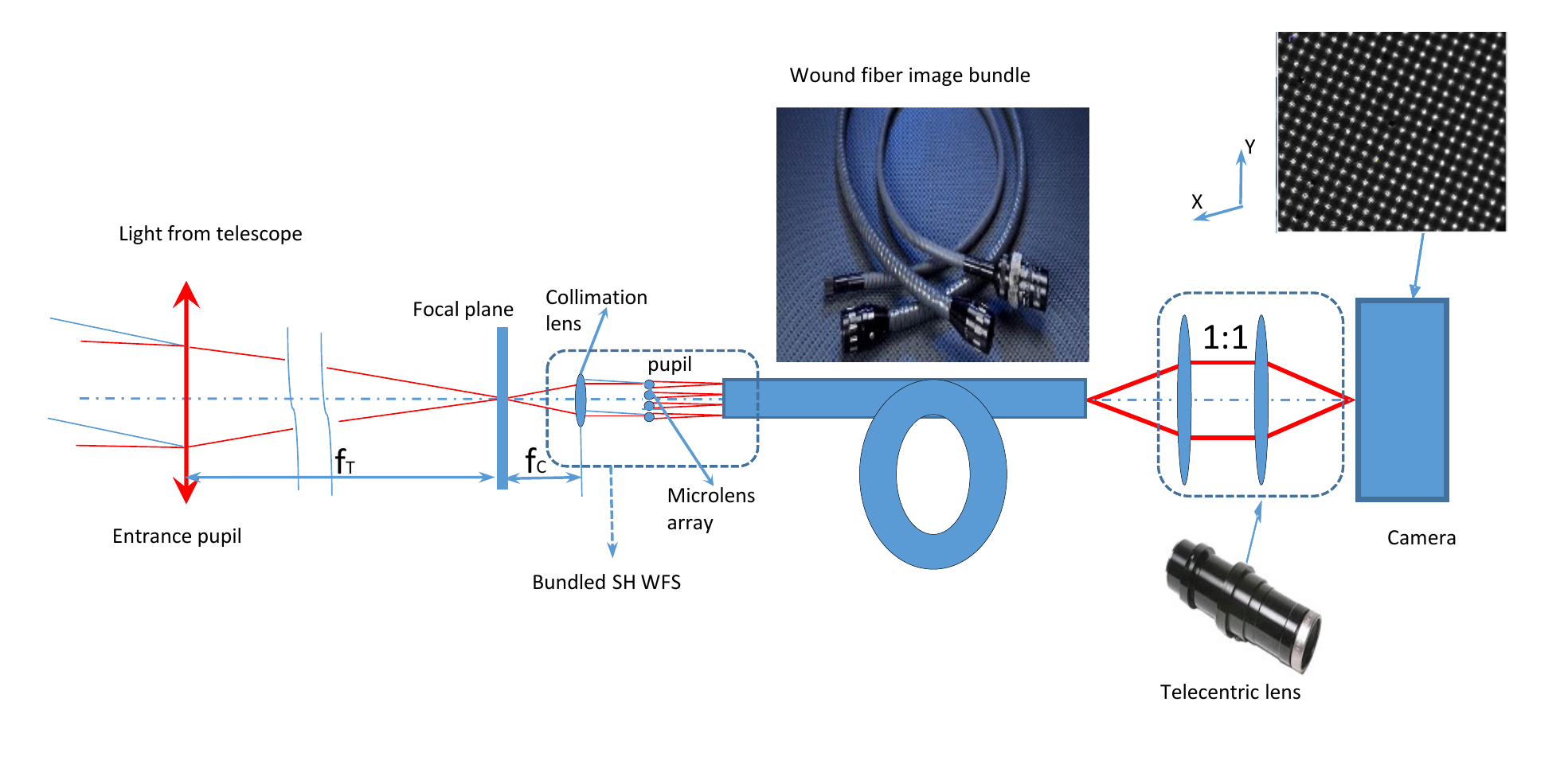}
		\caption{Shack-Hartmann wavefront sensor with coherent fibre image bundle }\label{SensorLayout}
	\end{center}
\end{figure*}

The measurement sensitivity of a conventional SHWFS mainly depends on the detector sampling, photon noise from the guide star, read-out noise of the camera, the speckle noise introduced by the atmosphere and algorithm to find the centroids(\cite{Irwan1999,Thomas2006MNRAS}). A detailed examination  of the different centroiding methods and image-processing techniques can be found in reference (\cite{Nightingale2013}).

While the wound fibre image bundle with high core density can relay the image effectively, its application in the SHWFS has to be examined carefully because of its unique properties. In this paper, we focus on the simulation of the Schott $'$demo$'$  wound fibre image bundle defects impact on the centroiding measurement accuracy. We report on the experimental measurements of this demo wound fibre image bundle using our test bench, configured as a SHWFS.

\section{CONCEPT OF THE SHACK-HARTMANN WAVEFRONT SENSOR WITH WOUND FIBER IMAGE BUNDLE}
\label{sec:Concept}

Figure \ref{SensorLayout} shows the details of Shack-Hartmann wavefront sensor\textcolor{black}{(SHWFS)} with the wound  fibre image bundle. \textcolor{black}{Note that the Shack-Hartmann wavefront sensor we discuss here actually includes a collimating lens in front of a microlens array.}The light from a telescope pupil is collimated \textcolor{black}{through the collimating lens and the telescope pupil is re-sized and re-imaged onto the microlens array}. The incoming light forms a diffraction limited spot array on the focal plane of the microlens array where the wound fibre image bundle is placed.  The spot array image is then relayed by the fibre image bundle. At the far end, the bundle surface (spot array) is then re-imaged onto the camera. By measuring the centroid of each spot in the spot array, the slope of the inbound wavefront can be found and hence the wavefront reconstructed. \textcolor{black}{Since the collimating lens is of small size and  good optical quality, the whole bundled SHWFS can be fit into the Starbug which is going to be located at the telescope focal plane.}

\begin{figure}
	\begin{center}
		\includegraphics[width= 1\columnwidth]{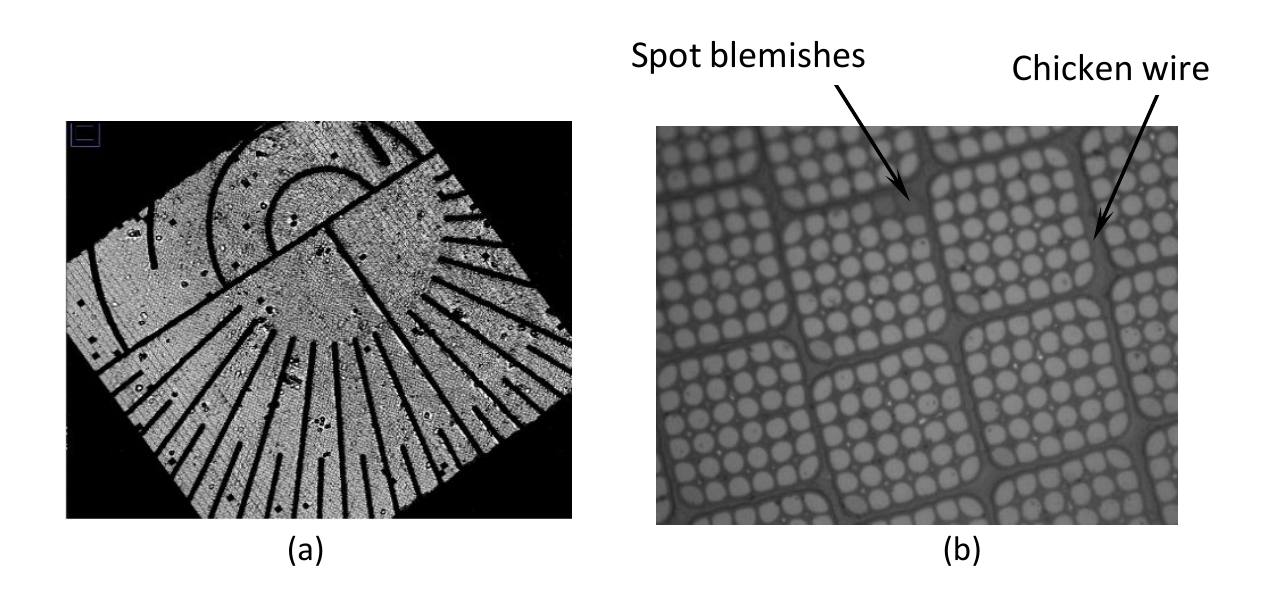}
		\caption{(a) Image of a target. (b) End image of the wound fibre image bundle with 40X magnification }\label{WoundBundle}
	\end{center}
\end{figure}

\section{PROPERTIES OF THE WOUND FIBER IMAGE BUNDLE} 

\begin{figure}
	\begin{center}
		\includegraphics[width= 1\columnwidth]{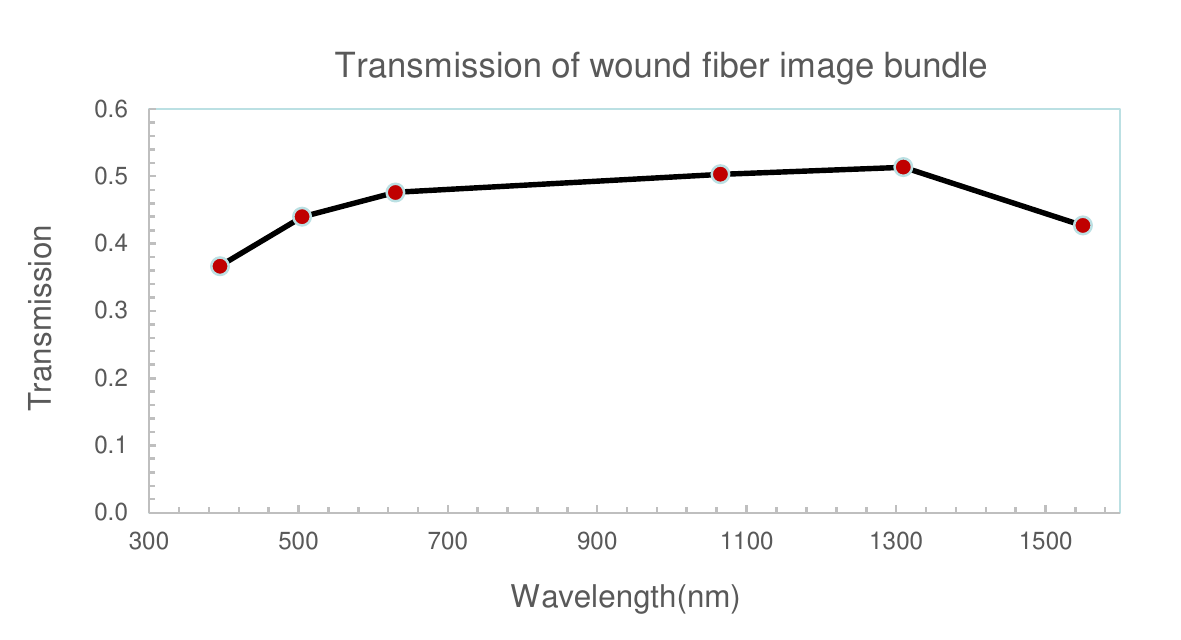}
		\caption{Measured transmission of wound fibre image fibre bundle }\label{Transmission}
	\end{center}
\end{figure}

The wound fibre image bundles normally comprise tens of thousands  optical fibres of either glass or polymer cores. Ideally, each fibre is considered to act as an independent waveguide. The wound fibre image bundle can be made with sizes up to tens of millimetres depending on the manufacturing. Because individual fibre positions at both ends of the bundle match, an image projected onto the input face is transmitted unaltered to the output end. This is demonstrated by Figure \ref{WoundBundle}(a) where a Thorlabs target is illuminated and imaged, with a magnification of 1 onto the input face of the wound fibre image bundle and at the output end, it is then re-imaged with a magnification of 2 onto the detector. 

The wound fibre image bundle used in our discussion is from Schott North America as a demonstration unit. It is 4mm by 4mm in square shape and 1m in length. Its numerical aperture is 0.6. The individual fibre size is 10$\mu$m with core size of 8$\mu$m. It is typically fabricated by winding a tile of  6$\times$6 multi fibre bundle cell into a single-layer ribbon on a cylindrical mandrel, assembling layers in a separate laminating operation, then cutting through and polishing the ends (\cite{Schott2007}). Its end surface is shown in Figure \ref{WoundBundle}(b). 

\subsection{Optical transmission}

To verify the operation of this demo wound fibre image bundle, its optical transmission was first measured. Figure \ref{Transmission} shows its  transmission for the wavelength range from 390nm to 1550nm with input beam F-number of 8. Note that it includes the surface Fresnel reflection from both end. Its transmission is above 40$\%$ for most visible and near infrared range. This implies that the demo wound fibre image bundle could  be used for wavefront sensing in the near infra-red. The loss mostly comes from the fill factor of the fibre image bundle which is the inherent property of image bundles.  

\begin{figure}
	\begin{center}
		\includegraphics[width= 0.9\columnwidth]{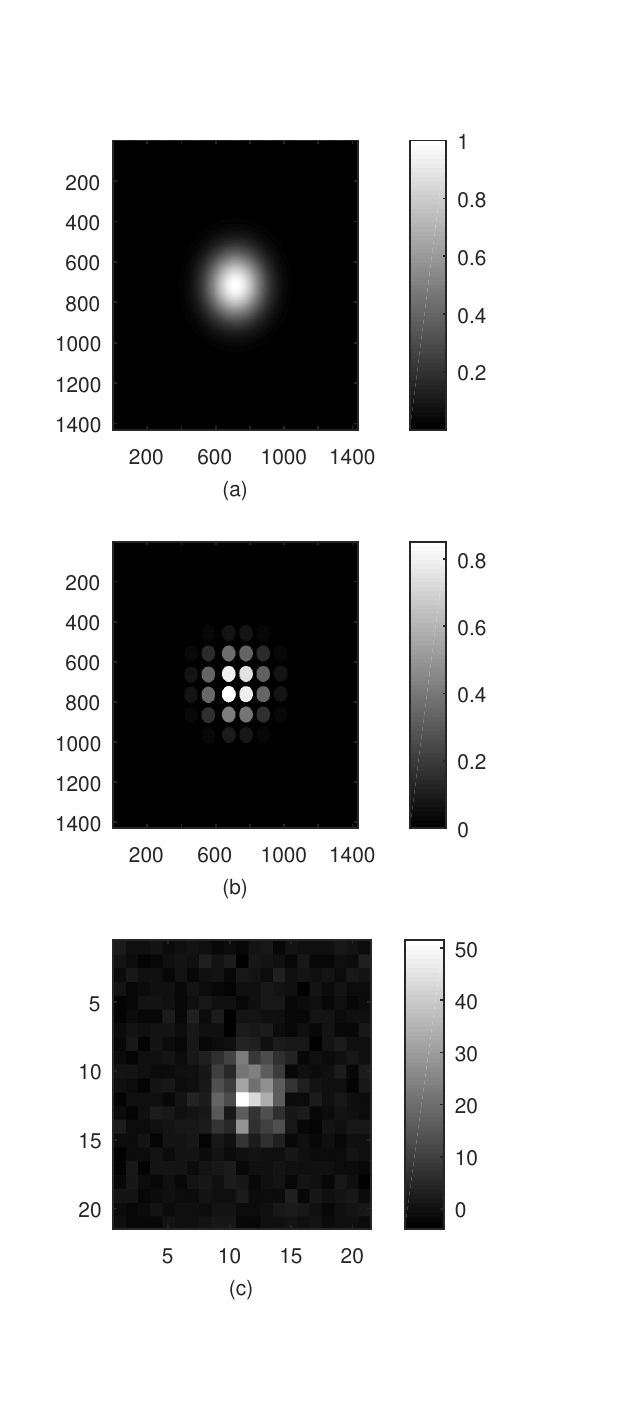}
		\caption{Simulated focused spot with equivalent  Airy size of 66.8$\mu$m on a 1430 $\times$ 1430 grid. (a)Pure Gaussian Beam. (b)Sampled by fibre image bundle. (c)Sampled by sCMOS chip(SNR:17) }\label{Gaussian}
	\end{center}
\end{figure}

\subsection{Defects}

The main two defects of the wound fibre image  bundle are blemishes and distortion(\cite{Schott2007}). The blemishes including spot and line blemishes ( also known as 'Chicken Wire') are shown in Figure \ref{WoundBundle}(b). The spot blemishes are the small areas (or groups of fibres)  with reduced or no transmission. The line blemishes are defined as a pattern of dark fibres that are two to four fibres wide at the fibre tile boundary. They are caused by damage to the fibres at the outside edges of the cell fibre array which can be caused by contamination, or improper temperature and pressure control in the pressing operation. Line blemishes are quantified by its length and quantity. The distortion could manifest itself in two forms: shear and gross distortion where the first one is defined as a lateral displacement that causes a straight line to be imaged as a $'$break$'$ line. It is caused by the misalignment of fibre tiles along the length. This results in a small break in the coherency of a relayed image. Gross distortion is defines as the distortion that causes a straight line to be imaged as a continuous curve. It is measured as the maximum displacement from a straight line and it ranges from less than 1$\%$ to 2$\%$ the clear aperture. 

\begin{figure}
	\begin{center}
		\includegraphics[width= 1\columnwidth]{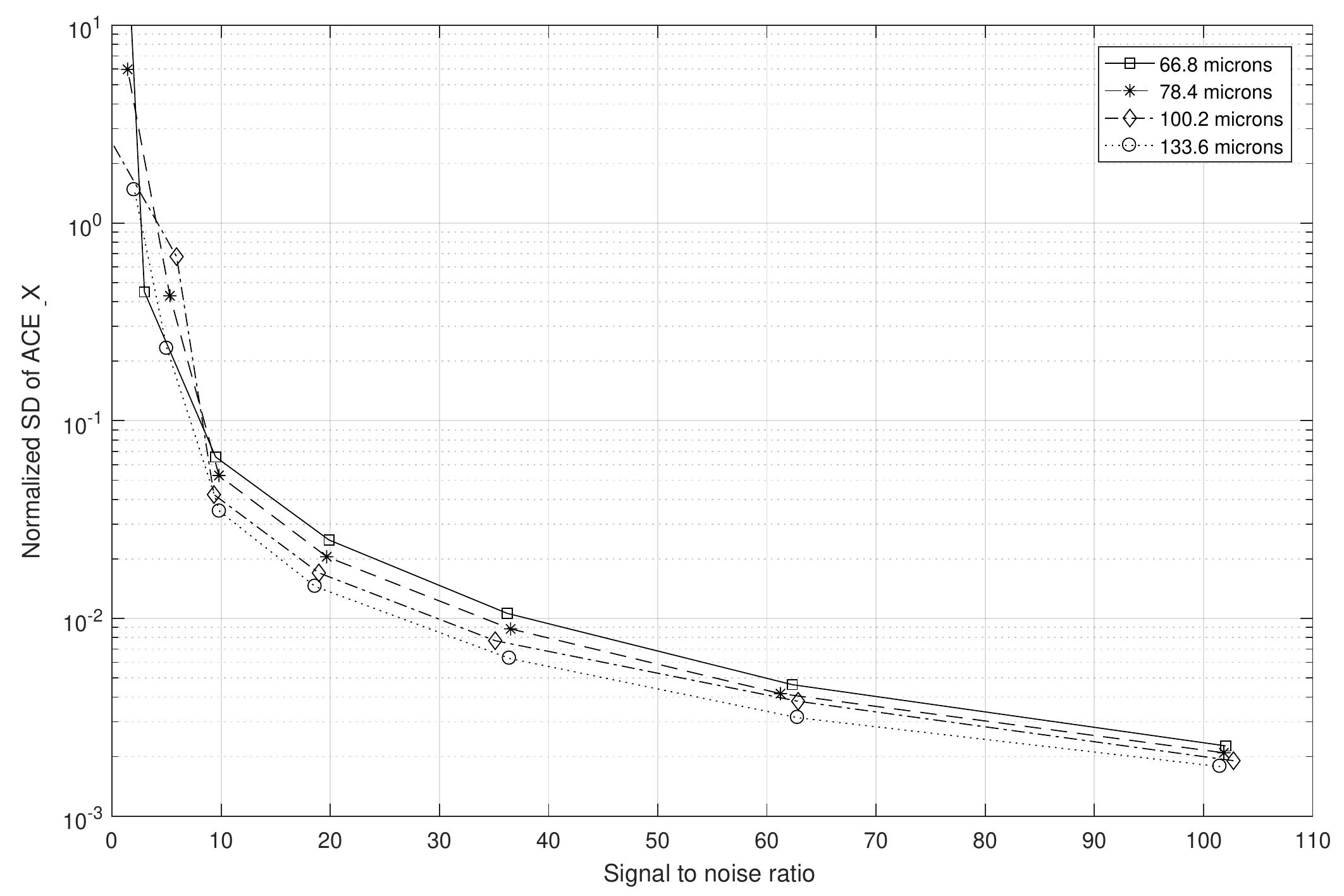}
		\caption{Measured normalized SD of ACE vs SNR with different spot sizes }\label{Result5}
	\end{center}
\end{figure}

\begin{figure*}
	\begin{center}
		\includegraphics[width= 1.7\columnwidth]{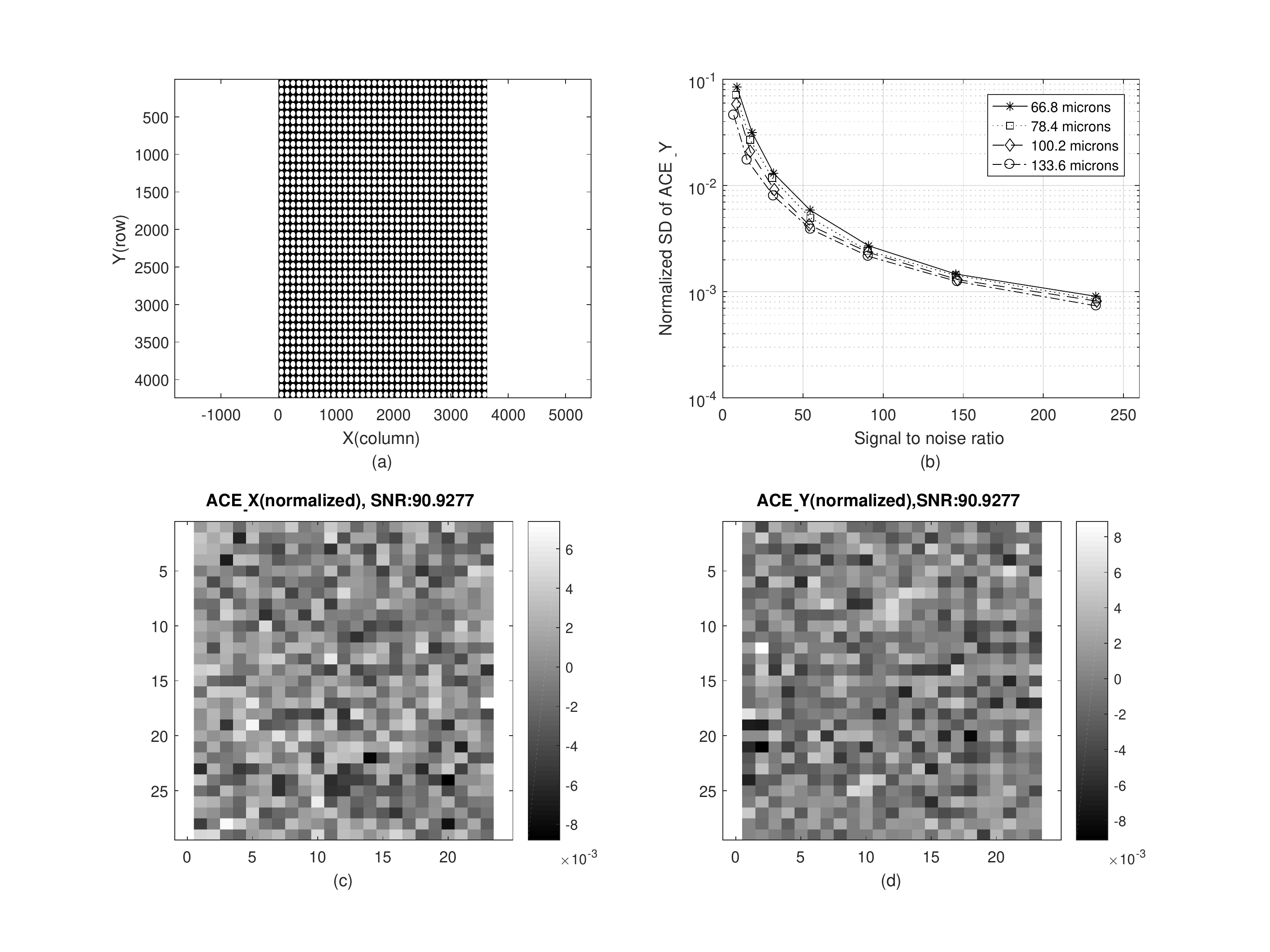}
		\caption{Simulation Results (a) simulated ideal fibre image bundle without any defects. (b) Normalize measured SD of ACE to the spot size vs different SNR. (c) Normalized measured $ACE\_X$. (d) Normalized measured $ACE\_Y$ when the SNR is 90.}\label{Result6}
	\end{center}
\end{figure*}

Since the spot array in the SHWFS are relayed through the fibre image bundle, the spots$'$ centroiding measurement accuracy would be influenced not only by the factors mentioned earlier as conventional SHWFS, but also exhibits its own unique properties. Various factors like fill factor, fibre image bundle defects, tile angle would make extra contributions to the centroiding measurement accuracy. Hence these centroid errors cause the reconfigured wavefront to be in error. It is therefore necessary to investigate the impact of those defects on the centroiding measurement accuracy.

\section{SIMULATION METHOD}
\label{Sec:Sim}

The Shack-Hartmann wavefront sensor simulation consists of a spatial resolution of 25$\times$25 microlens array (Thorlabs MLA150-7AR) which is available in our laboratory. Its focal length is 6.7mm with a lens pitch of 150$\mu$m. The simulated telescope aperture is 3.9m in diameter. The focused image is then sampled by the wound fibre image bundle and re-imaged at the other end with a 1$:$1 image relay system and captured by the camera. 

The accuracy of a reconstructed wavefront is largely dependent upon the determination of spot centre, commonly referred to the $'$centroiding$'$ of a focused spot. \textcolor{black}{It is the diffracted image of the sub-aperture of telescope formed by the lenslet of the microlens array.} The standard approach(the first moment calculation) is defined as: 
\begin{equation}
X_m = \dfrac{\sum_{i}^{n_i} \sum_{j}^{n_j} X_i *I{(i,j)}}{\sum_{i}^{n_i} \sum_{j}^{n_j} I{(i,j)}};
Y_m = \dfrac{\sum_{i}^{n_i} \sum_{j}^{n_j} Y_i *I{(i,j)}}{\sum_{i}^{n_i} \sum_{j}^{n_j} I{(i,j)}}
\label{equ:FirstMoment}
\end{equation}

where $X_{i}$ , $Y_{j}$ is  the  coordinate  position  of  the  (i, j)th   pixel  in one sub-aperture of the lenslet. $I{(i,j)}$ is the  input  wavefront  intensity  at the  pixel (i, j)  on the square sub-aperture  having  $n_{i} \times n_{j} $ pixels. It gives  a reasonable   pixel   accuracy   of   centroid   estimation   in   real  time calculation with minimal computational time and is widely adopted by adaptive optical community as the most efficient and robust method (\cite{Vyas2009}). We use this centroid method throughout the paper.

A Gaussian function defined as:
\begin{equation}
I(x,y) = exp(-0.5*(\frac{(X-X_{c})^2}{\sigma_{x}^2} + \frac{(Y-Y_{c})^2}{\sigma_{y}^2}))
\label{equ_Gaussian}
\end{equation}

where $X_c$ and $Y_c$ represents the true spot centre and $\sigma_{x}$ and $\sigma_{y}$ is the standard deviation of the Gaussian beam in two dimensions. To simplify analysis, these two are assumed the same.  Figure \ref{Gaussian}(a) shows the simulated Gaussian beam on a grid of 1430$\times$1430 pixels with the beam diameter of 66.8$\mu$m which is close to the airy disk of the focused spot created by each lenslet of the Shack-Hartmann wavefront sensor available in our lab. Figure \ref{Gaussian}(b) shows the focused spot sampled by the coherent fibre image bundle with each fibre of 101 pixels in diameter. Figure \ref{Gaussian}(c) shows the relayed and re-sampled spot on the sCMOS camera. Its typical area of interest (AOI) is about 25$\times$25 pixels on the sCMOS camera with pixel size of 6.5$\mu$m which is Zyla camera available in our lab(\cite{Andor2017}). \textcolor{black}{The AOI is determined by the microlens' pitch (the spots spacing). The spot size is chosen to represent the microlens array available for testing in our bench-top laboratory setup. The same methodology  can be applied to any configuration of wound fibre bundle SHWFS}. 

The signal to noise ratio (SNR) of the measurement within the AOI can be calculated as:

\begin{equation}
\frac{S}{N} = \frac{S*\sqrt{t}}{\sqrt{S + n_{pixel} * (B + I_{d} + \frac{R^2}{t})}}
\end{equation}

Where t is the integration time per exposure, $n_{pixel}$ is the number of pixels in the AOI, S is the signal received within one lenslet and B is the sky background, Id and R are dark current and read noise of the Andor Zyla sCMOS camera respectively.

To compare the centroiding accuracy at different simulation conditions quantitatively, the absolute centroiding measurement error $($ACE$)$  is defined as:

\begin{align*}
ACE_-X &= X_{m} - X_{c}\\
ACE_-Y &= Y_{m} - Y_{c}\\
ACE_-R &= \sqrt{(X_{m} - X_{c})^2 + (Y_{m} - Y_{c})^2 }
\end{align*}

Where ($X_{m}$,$Y_{m}$) is the estimated spot$'$s centre on the detector focal plane, ($X_{c}$,$Y_{c}$) is the simulated spot$'$s  true centre. 

The simulation process is:
\begin{itemize}
	\item Generate a Gaussian beam with a  known centre.
	\item Generate the wound fibre image bundle mask with tile angle randomly chosen between 0 and specified angle with chicken wire row and column width chosen randomly.
	\item The Gaussian beam is multiplied with the fibre image bundle mask.
	\item Take the mean intensity of each fibre within the AOI.
	\item Pixelation (binning) of the image with the detector and add read noise, \textcolor{black}{dark current,and photon noise by using the specification sheet from Andor} to the  AOI.
	\item Scan the Gaussian beam  along fibre image bundle in two dimensions.	
	\item Calculate the ACE at each position.
	\item Repeat the same process with a different sizes of the Gaussian beam.
\end{itemize}

We then normalize the standard deviation(SD) of ACE at each simulation condition by the Airy beam size which is 5.81$\sigma$, where $\sigma$ is defined in equation \ref{equ_Gaussian}. 
 
\section{SIMULATION RESULTS}
 
\begin{figure*}
	\begin{center}
		\includegraphics[width= 1.7\columnwidth]{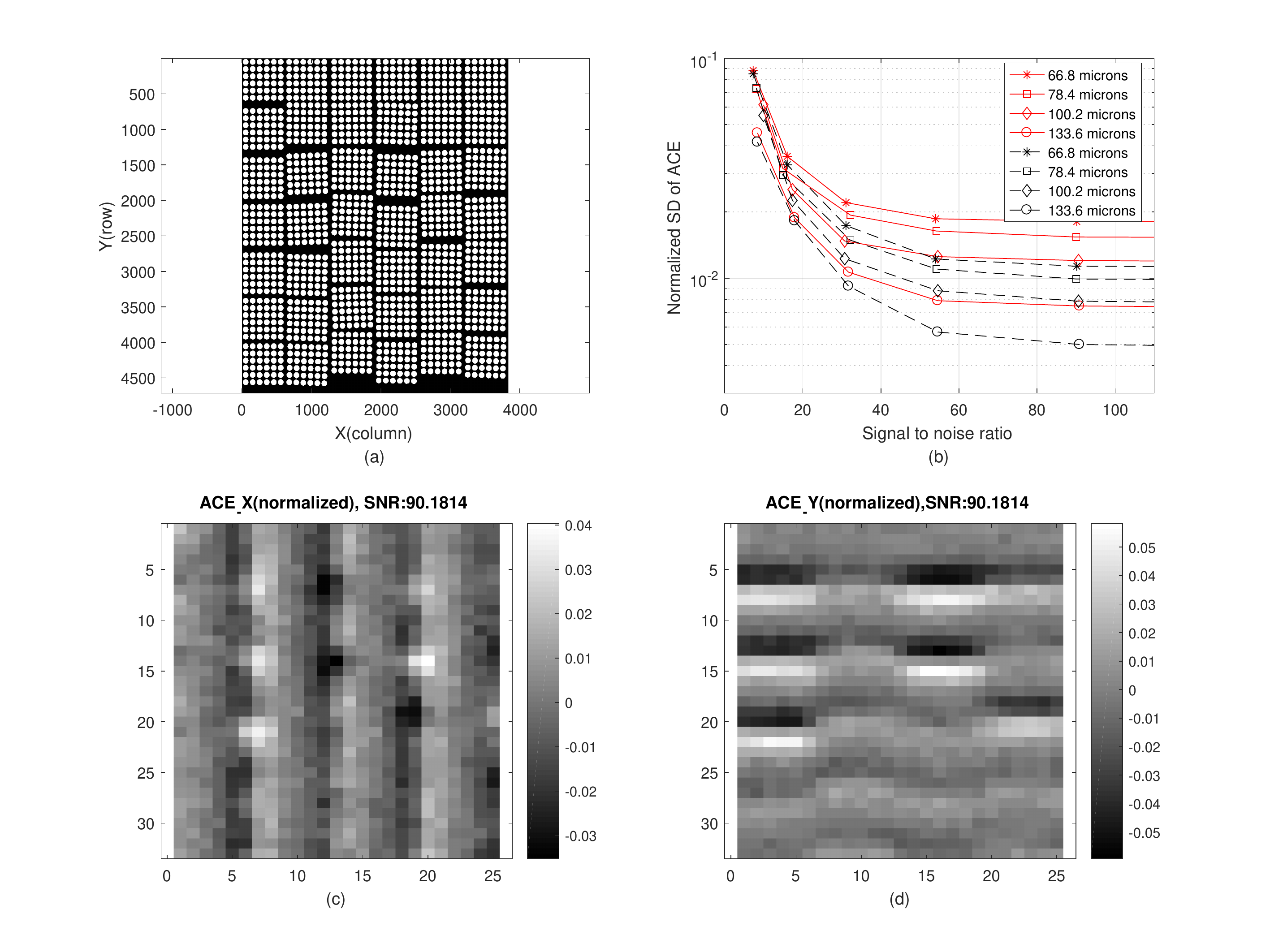}
		\caption{Simulation Results. (a)Simulated Fibre bundle with tile angle uniform random with range -2 to 2 degree and chicken wire width uniform random in row is 0 to 2$\mu$m and in column is 0 to 10$\mu$m . (b)Normalized SD of ACE vs. SNR with different spot size. The solid lines are for ACE in row. The dashed lines are for ACE in column. (c)Normalized measured $ACE\_X$. (d)Normalized measured $ACE\_Y$. with SNR of 102.} \label{Result7}
	\end{center}
\end{figure*}

\begin{figure}
	\begin{center}
		\includegraphics[width= 1\columnwidth]{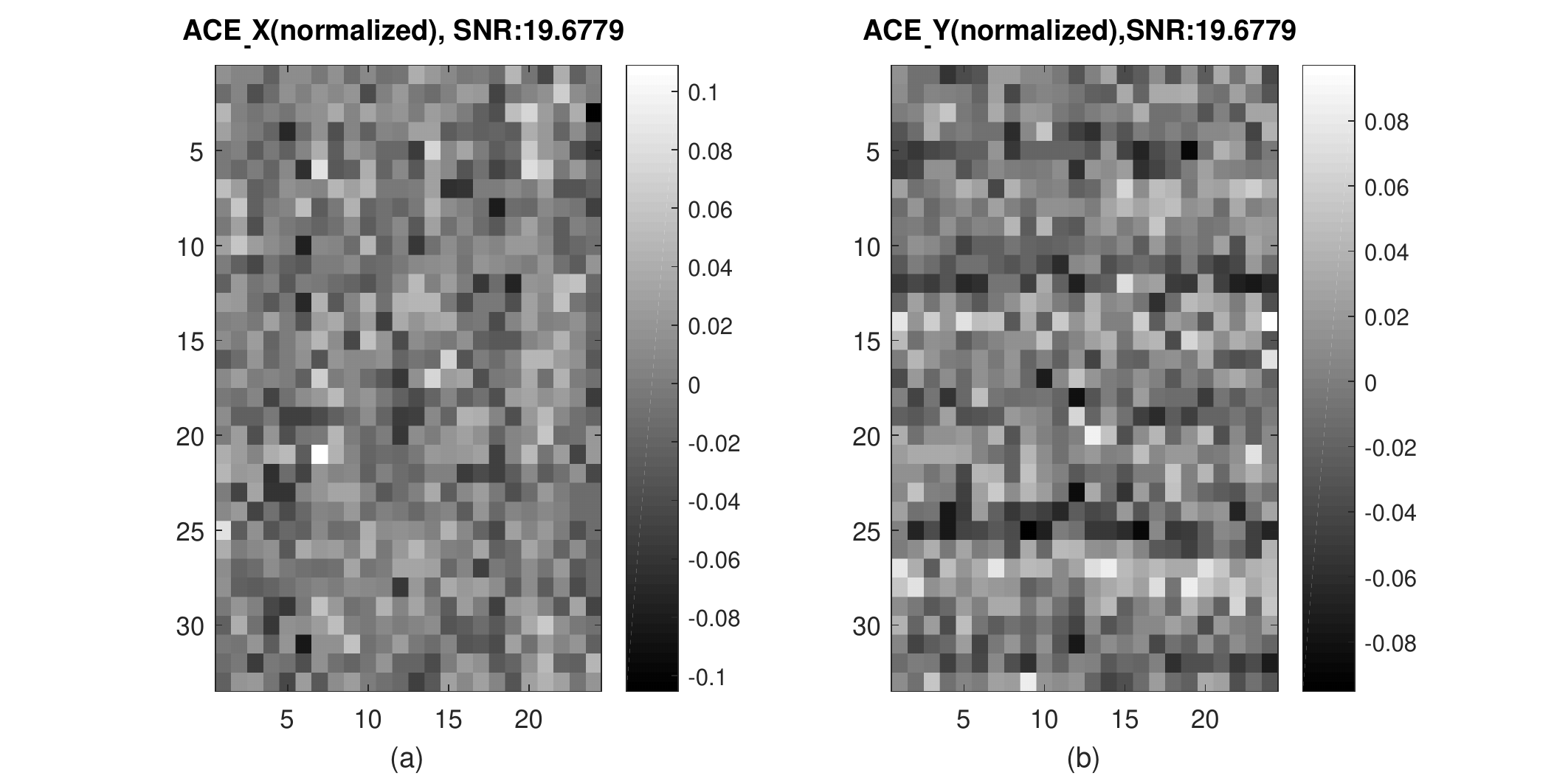}
		\caption{(a)Normalized measured $ACE\_X$ 
When SNR is 19. (d)Normalized measured $ACE\_Y$ When SNR is 19.}\label{Result8}
	\end{center}
\end{figure}

\begin{figure}
	\begin{center}
		\includegraphics[width= 1\columnwidth]{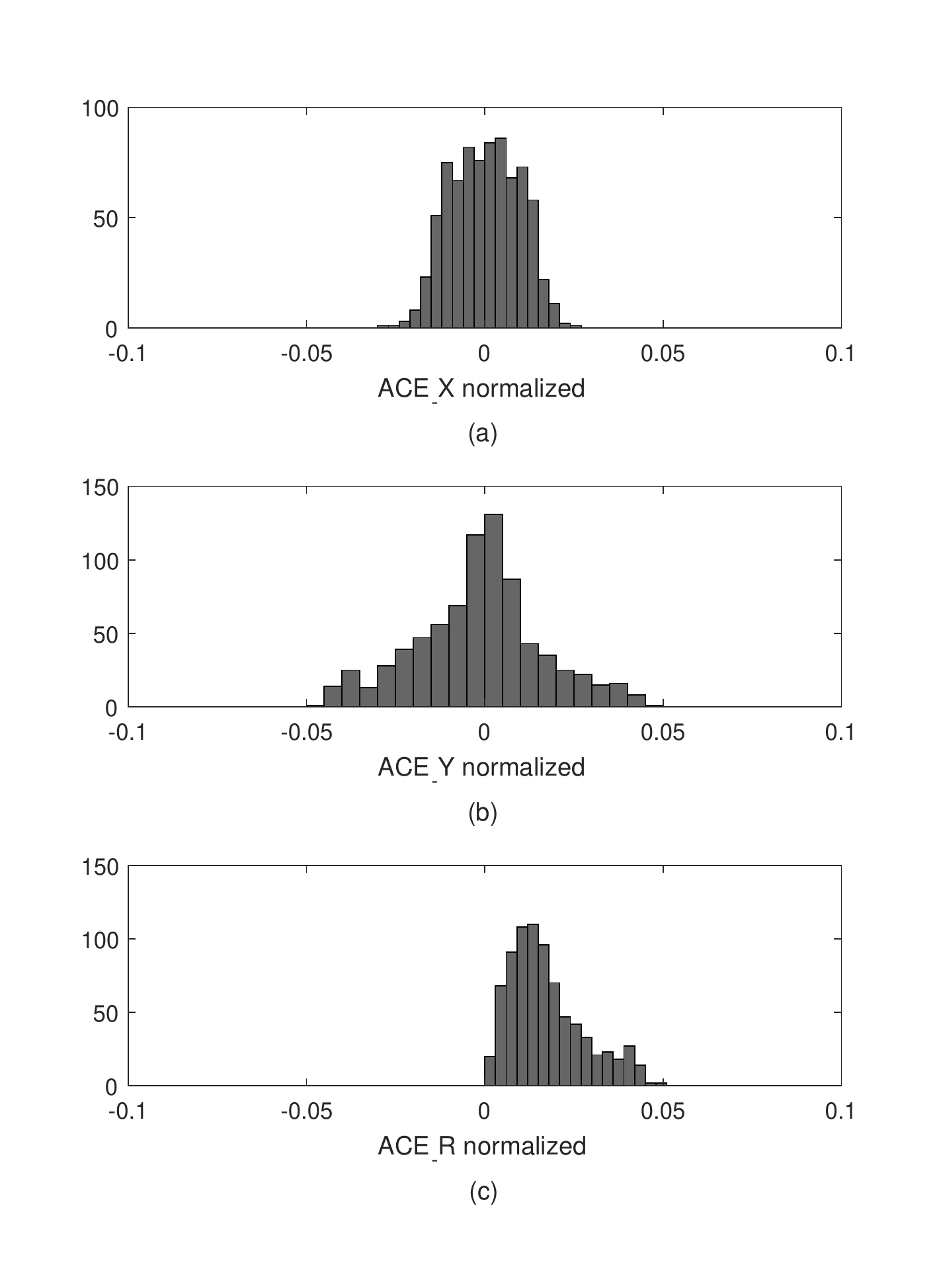}
		\caption{Histogram of ACE for the Figure \ref{Result7}(c),(d) data set in column(top), row (middle) and R direction(bottom).}\label{Result9}
	\end{center}
\end{figure}

\begin{figure}
	\begin{center}
		\includegraphics[width= 1\columnwidth]{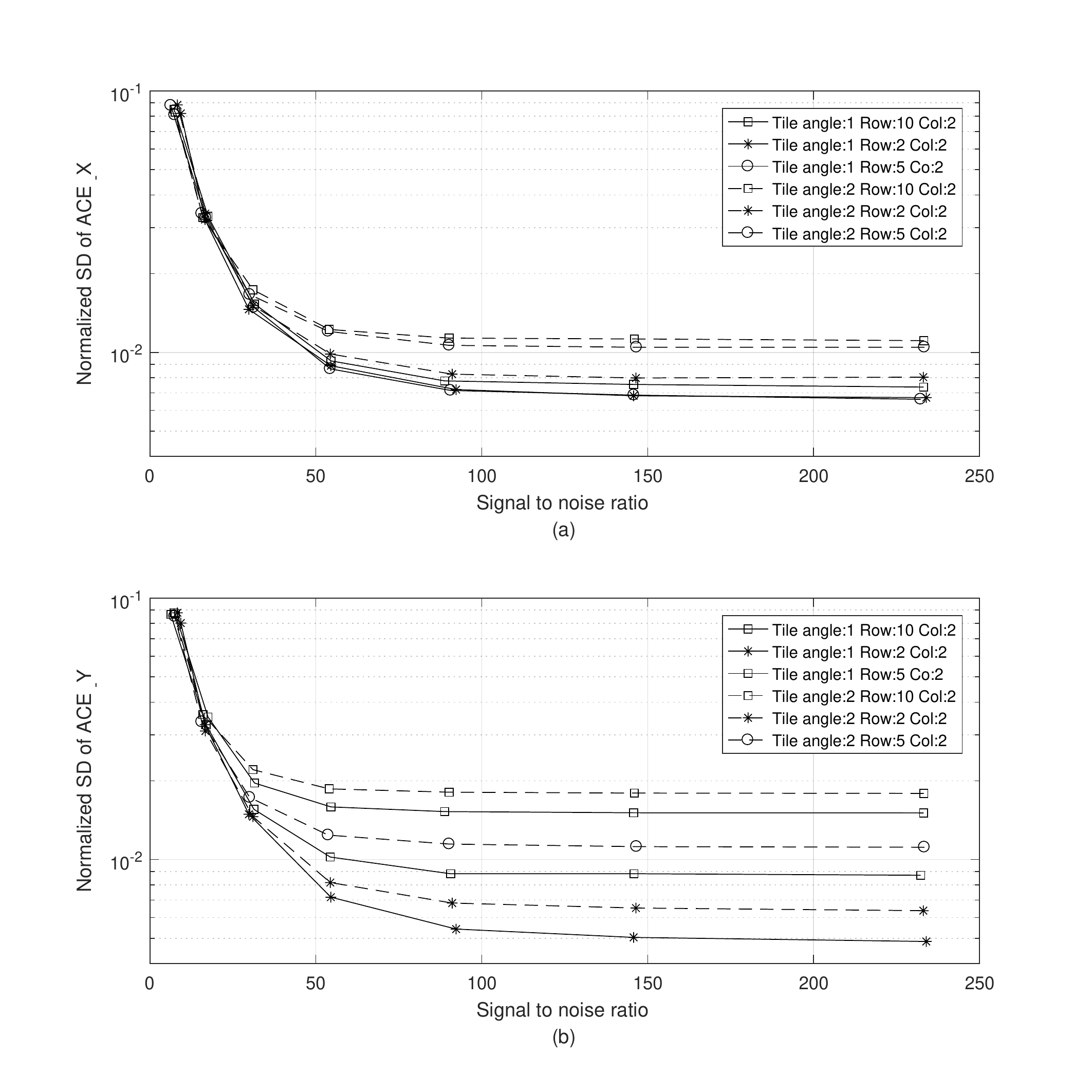}
		\caption{Measured normalized SD of ACE vs SNR with spot size of 66.8$\mu$m. (a) in column . (b) in row.}\label{Result10}
	\end{center}
\end{figure}

Since the available fibre core size of the wound fibre image bundle is fixed by the manufacturer, we won$'$t simulate the fibre core size impact on the performance of the SHWFS. A fibre diameter of 10$\mu$m with core size of 8$\mu$m is used for all our simulations.

The simulation is done by applying different magnitude of the star received on the wavefront sensor. The measured signal to noise ratio (SNR) range from 1 to 100 can be obtained either by increasing the guide star magnitude or the exposure time of the sensor. However in practice, increasing exposure time is very limited as it would exceed the atmospheric coherence time. For the simulations with SNRs above 80 (for an AOI of 23x23 pixels), it indicates low noise level and  SNR lower than 30 represents high noise level or poor sensor imaging. At SNR less than 10, the spot becomes virtually indistinguishable.

\subsection{Centroiding accuracy baseline of a conventional SHWFS}

Figure \ref{Result5} shows the simulated normalized standard deviation (SD) of the ACE vs SNR when the focused spot size is directly received by sCMOS camera. The spot size shown in Figure \ref{Result5} is the equivalent Airy disc size of the Gaussian beam. Note that only ACE for column is shown here as ACE for row is the same. This simulation provides the centroiding accuracy baseline for a conventional SHWFS. Spot size which is in the range from 66.8$\mu$m to 133.8$\mu$m are studied. \textcolor{black}{The variation of the spot size is to assess the sampling performance of the wound bundle SHWFS}.  It is shown that the ACE decreases when the SNR is increases for the different sizes of spots. Also,the ACE experiences only small decreases as the spot size gets larger. This is expected as under this condition, the measurement error is only related to the camera pixelation error and its read  noise. To improve the centroiding accuracy, it is most effective to improve the measurement SNR.

\subsection{SHWFS with an ideal fibre image bundle}
Figure \ref{Result6} shows the simulation results  when an ideal fibre image bundle is constructed without any defects which is similar to the case for the polymer image fibre bundle. Figure \ref{Result6}(a) shows the image of the  simulated fibre image bundle.  Figure \ref{Result6}(b) shows the measured centroiding accuracy at row direction vs SNR with difference spot size. The performance is very similar to the previous simulation where there is no coherent fibre image bundle.  Figure \ref{Result6}(c) and (d) show that the normalized ACE when the spot (66.8$\mu$m) scanning through row and column direction respectively. It is shown that there are no features within the ACE measurements. Note that the peak to valley measurement centroiding error in both direction are very similar. For low SNR $<$10, the ACE increases noticeably for smaller spot sizes. We note the similarity in the ACE between the conventional SHWFS (Figure \ref{Result5}) and that of the ideal fibre image bundle ( Figure \ref{Result6}(b)), as the spot is well sampled by the detector.

\subsection{Wound fibre image bundle with cell's tile angle and chicken wire defects}

Figure \ref{Result7}(a) shows the simulated wound fibre image bundle with the tile angle uniformly random distributed from  $-2^{o}$ to $2^{o}$. The chicken wire width uniformly random in column is 0 to 2$\mu$m and in row is 0 to 10$\mu$m respectively. Figure \ref{Result7}(b) shows that the measured normalized SD of ACE vs SNR with different spot size at both row and column direction. It is shown that the normalized SD of ACE in row is bigger than it in column direction. This is because the chicken wire mean gap in row is 5 times bigger than the column mean gap. It is also shown that the SD of ACE reduces when the SNR is increasing. The normalized SD of ACE is also reduced when the spot size increases. This is because that when the spot is small, its focused spot would be sampled by less fibres, hence the line blemishes (chicken wire) impact more on the centroid measurement. However, as the spot size increases, the coverage of fibres average out the bundle$'$s defect effects and hence the measured normalized SD of ACE is less sensitive. Figure \ref{Result7}(c) shows the measured normalized ACE when the spot (66.8$\mu$m) scanning over the column direction. It clearly shows that the measurement error is fluctuated with the chicken wire position in column.  Figure \ref{Result7}(d) shows the measured normalized centroid error when the spot (66.8$\mu$m) scanning over the row direction. It is also shown that when the SNR is getting bigger (i.e. SNR $>$ 50), the chicken wire effect becomes obvious. When the SNR is smaller, the chicken wire effects are less important to the ACE since the read noise become the dominant factor. 

 \begin{figure*}
	\begin{center}
		\includegraphics[width= 1.7\columnwidth]{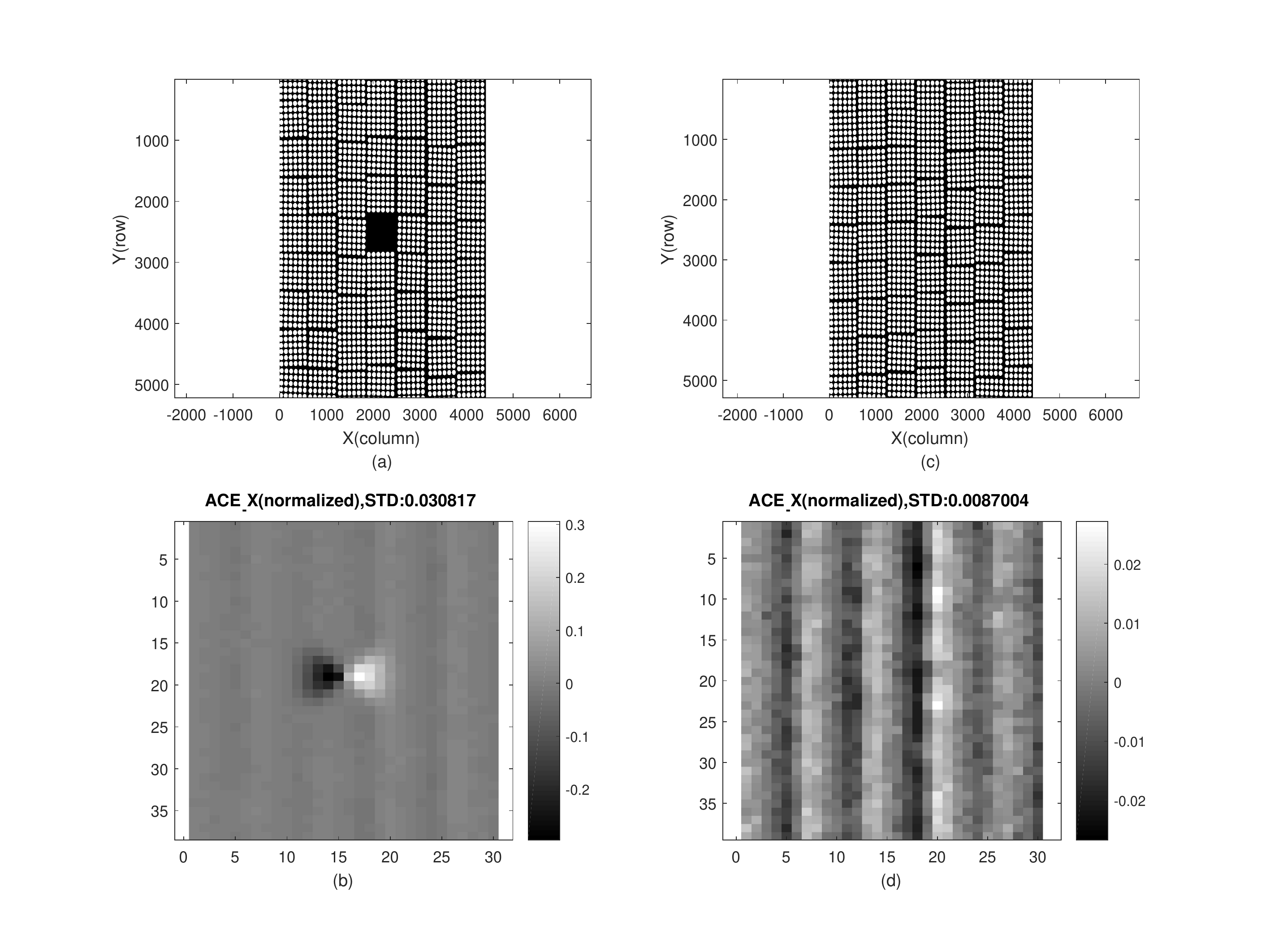}
		\caption{(a) Simulated fibre image bundle with blemish and chicken wire. (b) Simulated fibre image bundle with only chicken wire. (c)Normalized measured $ACE\_X$ for simulated bundle(a) . (d)Normalized measured $ACE\_X$ for simulated bundle(b). with SNR of 90.} 
        \label{Result11}
	\end{center}
\end{figure*}

Figure \ref{Result8} shows that with SNR of 19 and same fibre image bundle, the measured normalized ACE in both column and row direction. It is obvious that now the centroiding error is not influenced by the chicken wire gap or bundle defects. 

Figure \ref{Result9} shows the histogram derived from Figure \ref{Result7} $(c),(d)$ for spot size of 66.8$\mu$m. The standard deviation of the measured ACE, normalised by the spot size, in column is about 0.0115 while in row direction it is about 0.0181. 

Figure \ref{Result10} shows the simulated normalized SD of ACE vs SNR  with different fibre image bundle parameters. It is shown that with the focused spot size of 66.8$\mu$m,  the tile$'$ angle uniformly random distributed -1 to 1 degrees and -2 to 2 degrees for the fibre wound bundle respectively. The normalized SD of ACE is decreasing with the SNR increases, however, it is not changing that significantly between these two angle distributions. For the same tile angle distribution, when the chicken wire width changes in row from distribution 0 to 2$\mu$m to the distribution 0 to 10$\mu$m, while the width of column keeps the same distribution 0 to 2$\mu$m, the centroiding accuracy in row direction is noticeably worse. Note that when the SNR over 50, increasing SNR won$'$t improve the centroiding accuracy significantly. 
 
 \subsection{Comparison of wound fibre image bundle with spot blemish}
 
In the previous section, we simulated the influence of  chicken wire width effects to the accuracy of the measured ACE. In this section, we discuss the spot blemish effects on the accuracy of the measured of ACE.  Figure \ref{Result11}(a),(b) shows the simulated wound fibre image bundle with and without spot blemish respectively. The two wound bundles are with similar chicken wire width of 2$\mu$m in both direction and tile angle of 2 degree simulated the same as previous section.  Figure \ref{Result11}(c),(d) shows the measured SD of the ACE corresponding to the two wound bundles. Note that the SNR for both simulation condition is 102. The measured standard deviation of ACE for wound bundle with spot blemish is 0.03 while for bundle without blemish it is 0.009. It is obvious that the measured SD of the ACE increases quite significantly when there is the spot blemish within the wound fibre bundle.

\section{EXPERIMENTS}

\begin{figure}
	\begin{center}
		\includegraphics[width= 1\columnwidth]{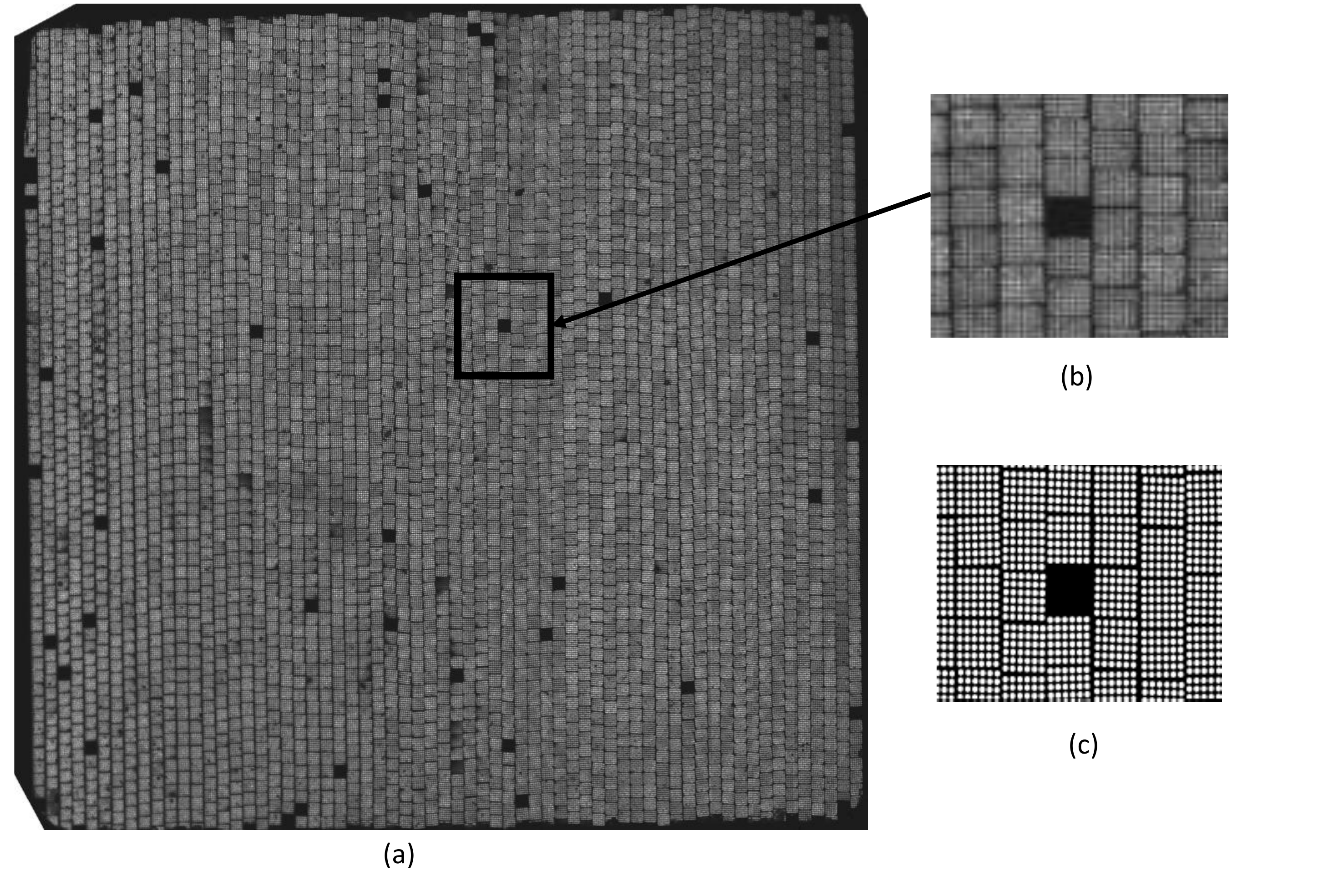}
		\caption{(a)Demo wound bundle used in experimental evaluation.(b)Magnified Section of the demo wound bundle.(c)Simulated wound bundle with blemish. }
		\label{WoundBundleSurface}
	\end{center}
\end{figure}

In this section we present the experimental evaluation of a demo wound fibre bundle from Schott North America. The demo bundle specifications are described in Section \ref{sec:Concept}. The imaging surface is shown in Figure \ref{WoundBundleSurface}. We see that the transmission of the fibres in Figure \ref{WoundBundleSurface} are relatively uniform and hence suitable for the application of Shack-Hartmann wavefront sensing. We therefore evaluate several aspects of the demo wound bundle: (1) blemishes - non-transmitting tiles; (2) distortion - misaligned tiles and (3) wavefront measurement.   

\subsection{Blemishes - Non-transmitting tiles}

The demo wound fibre bundle has a number of non-transmitting tiles (blemishes) observed as dark squares in Figure \ref{WoundBundleSurface} and zoom shown in Figure \ref{Misalignment}. We measured about 35 full and 12 partial non-transmitting tiles. The total non-transmitting tiles equate to about 1$\%$ of the imaging surface area. The Schott commercial wound bundle specifications for a comparable size, state a much lower defect count, less than 6 non-transmitting tiles. Non-transmitting tiles could potentially be avoided during the alignment of the microlens array to the bundle imaging surface. The spots that do coincide with a non-transmitting tile (few in number) can be excluded from the wavefront reconstruction. Alternatively, the implementation of calibration algorithms for those spots might suffice.     

\subsection{Distortion - Misaligned tiles}

\begin{figure}
	\begin{center}
		\includegraphics[width= 1\columnwidth]{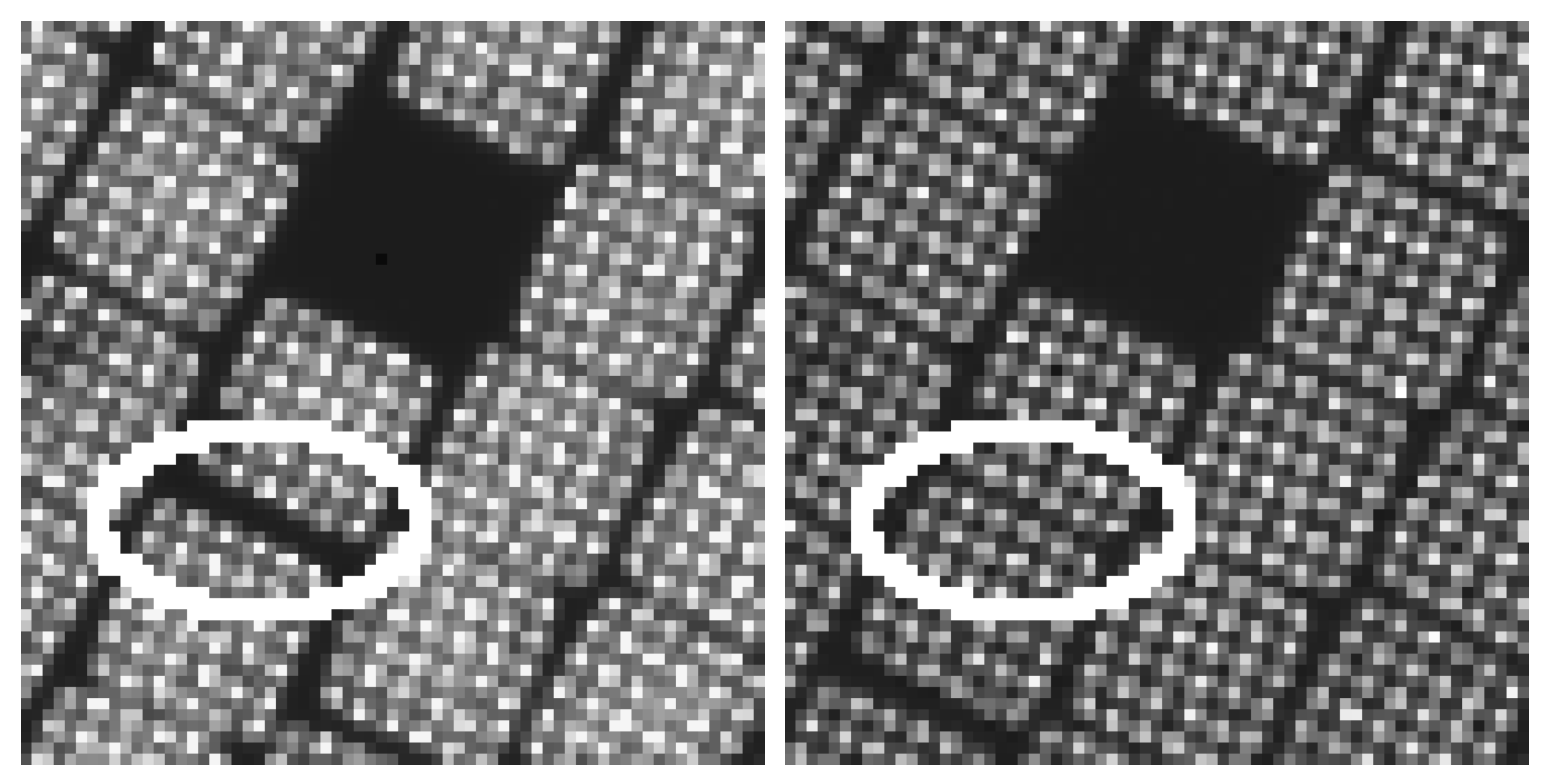}
		\caption{A misaligned tile (relative to surrounding tiles) found by comparing images of the front surface (left) and back surface (right). A non-transmitting tile is also shown at image top. }\label{Misalignment}
\end{center}
\end{figure}

We define misaligned tiles (distortion) as tiles that are displaced relative to surrounding tiles found by comparing the front and back surfaces of the wound bundle. An example of a misaligned tile is shown in Figure \ref{Misalignment}. A misaligned tile can be displaced as much as a fibre core (10$\mu$m)  and can result in an error the Shack-Hartmann spot location. The misaligned tiles can also occur between columns where the entire column can have a slight vertical height offset. Individual misaligned tiles, an example in Figure \ref{Misalignment} are less in number (only few observed) than the non-transmitting tiles. A misaligned tile is less of an issue as the Shack-Hartmann wavefront sensor can be nulled to the reference wavefront. The Shack-Hartmann wavefront sensor is primarily concerned with the spot displacement from its reference position.  

\subsection{Wavefront measurement}

\begin{figure}
	\begin{center}
		\includegraphics[width= 1\columnwidth]{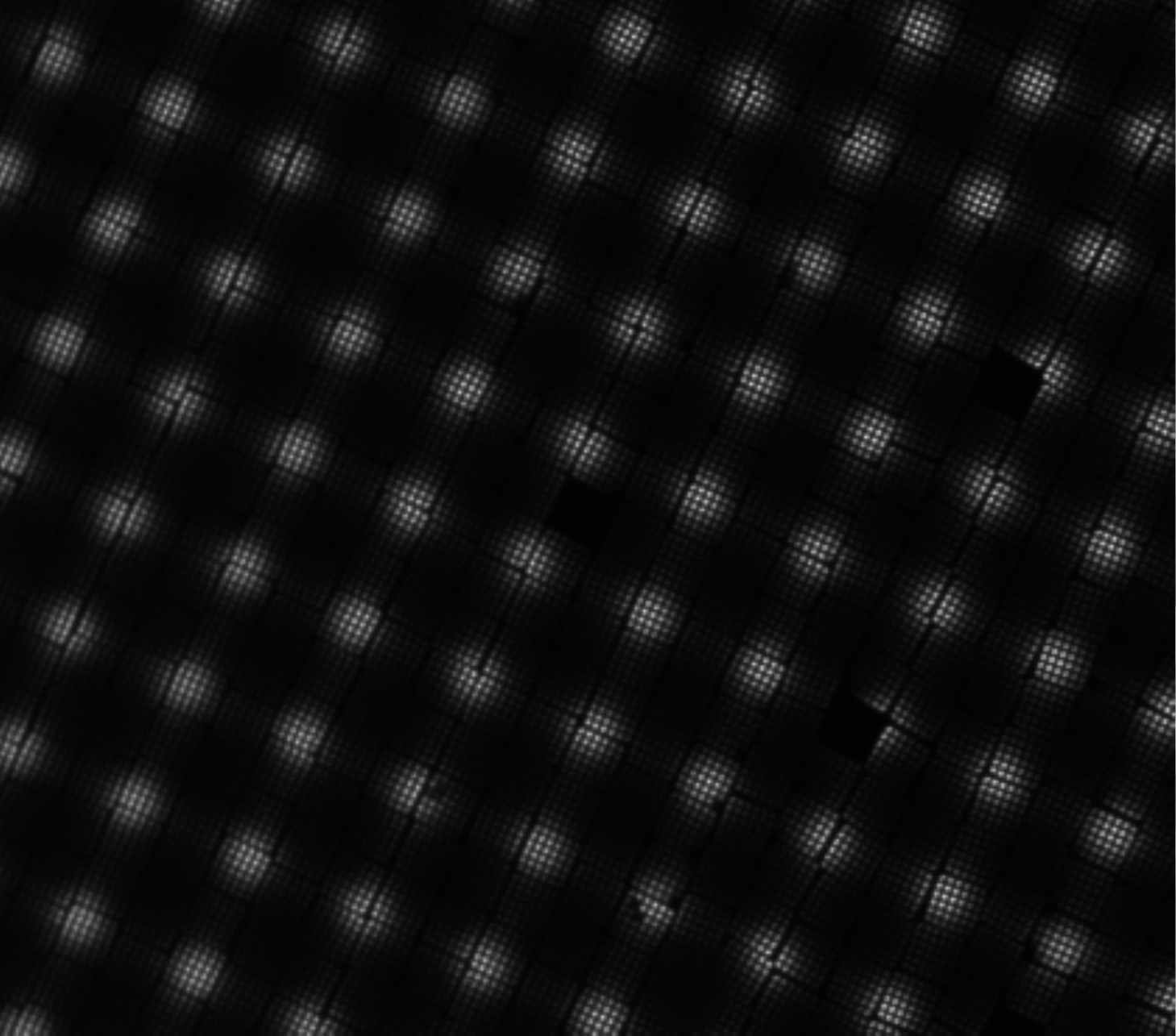}
		\caption{Zoom of Shack-Hartmann spot array as imaged by the demo wound bundle. Each sampled spot shows the resolved fibres and tile structure. }\label{WFS1}
\end{center}
\end{figure}

To examine how the non-transmitting and misaligned tiles impact a wound bundle based Shack-Hartmann wavefront sensor, we investigate the wavefront measurement residuals for different displacements of the spot array. The unique sampling of the spot array and the resulting wavefront measurement can be compared. The difference between the tip-tilt subtracted wavefronts can then be attributed to the bundle sampling effects of the non-transmitting and misaligned tiles. An example image of the Shack-Hartmann spot array is shown in Figure \ref{WFS1}.    

The Shack-Hartmann array is formed by using a microlens array (Thorlabs MLA150-7AR) having a lens pitch of 150$\mu$m and focal length of 6.7mm, focused onto the front surface of the wound fibre bundle. The bundle back surface is re-imaged onto an Andor Zyla 5.5 camera with a pixel size 6.5$\mu$m using a 2x telecentric lens (Edmund Optics \#58-431). The spot centroid calculated using the first moment calculation as described in Section \ref{Sec:Sim}. The spot diameters formed by the microlens array are approx. 71$\mu$m given by ${2.44\lambda N}$, where $\lambda$ is wavelength of 632nm and $N$ is the f-number $(f / \#)$.

\begin{figure}
	\begin{center}
		\includegraphics[width= 0.45\columnwidth]{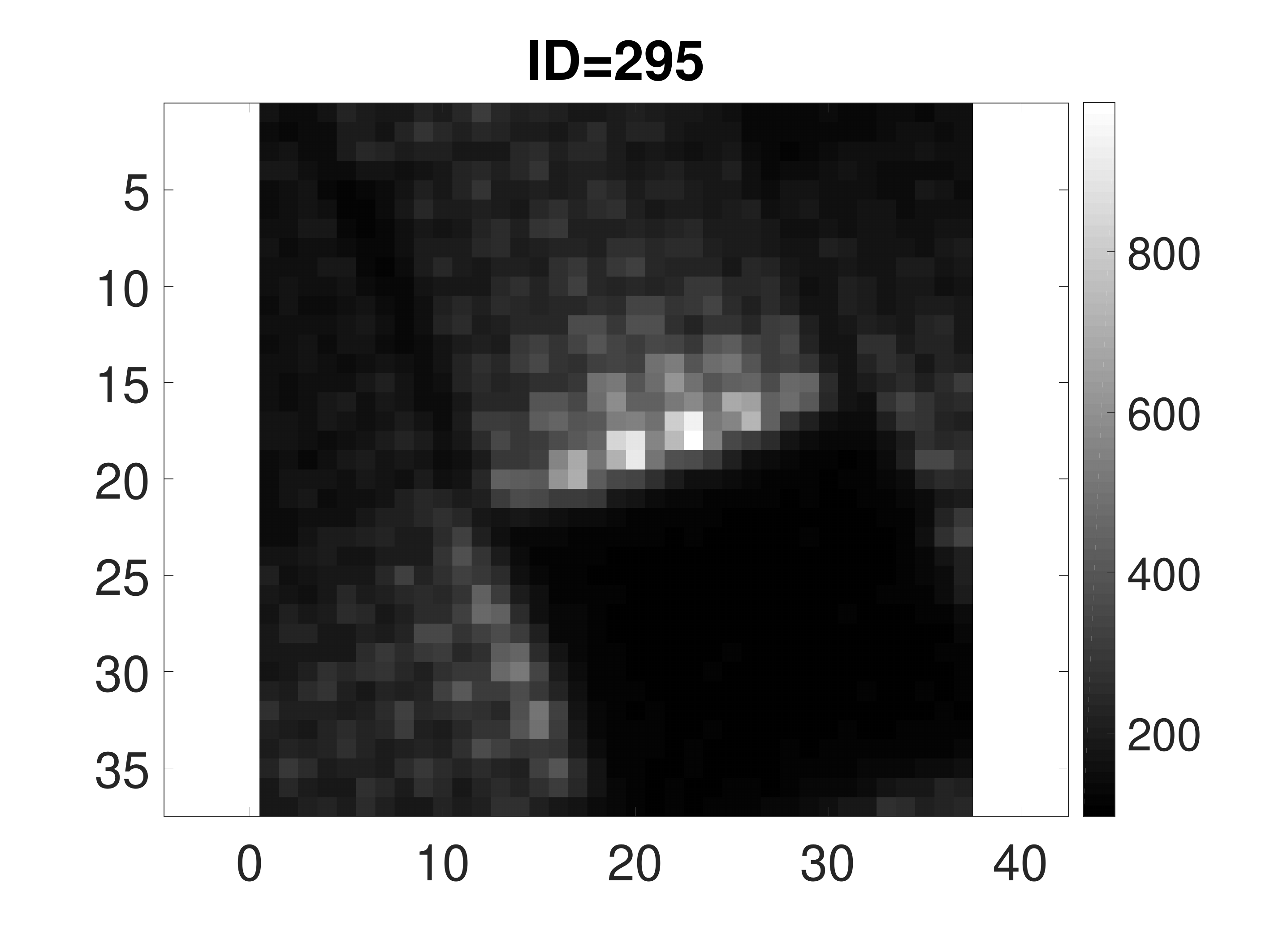}
		\includegraphics[width= 0.45\columnwidth]{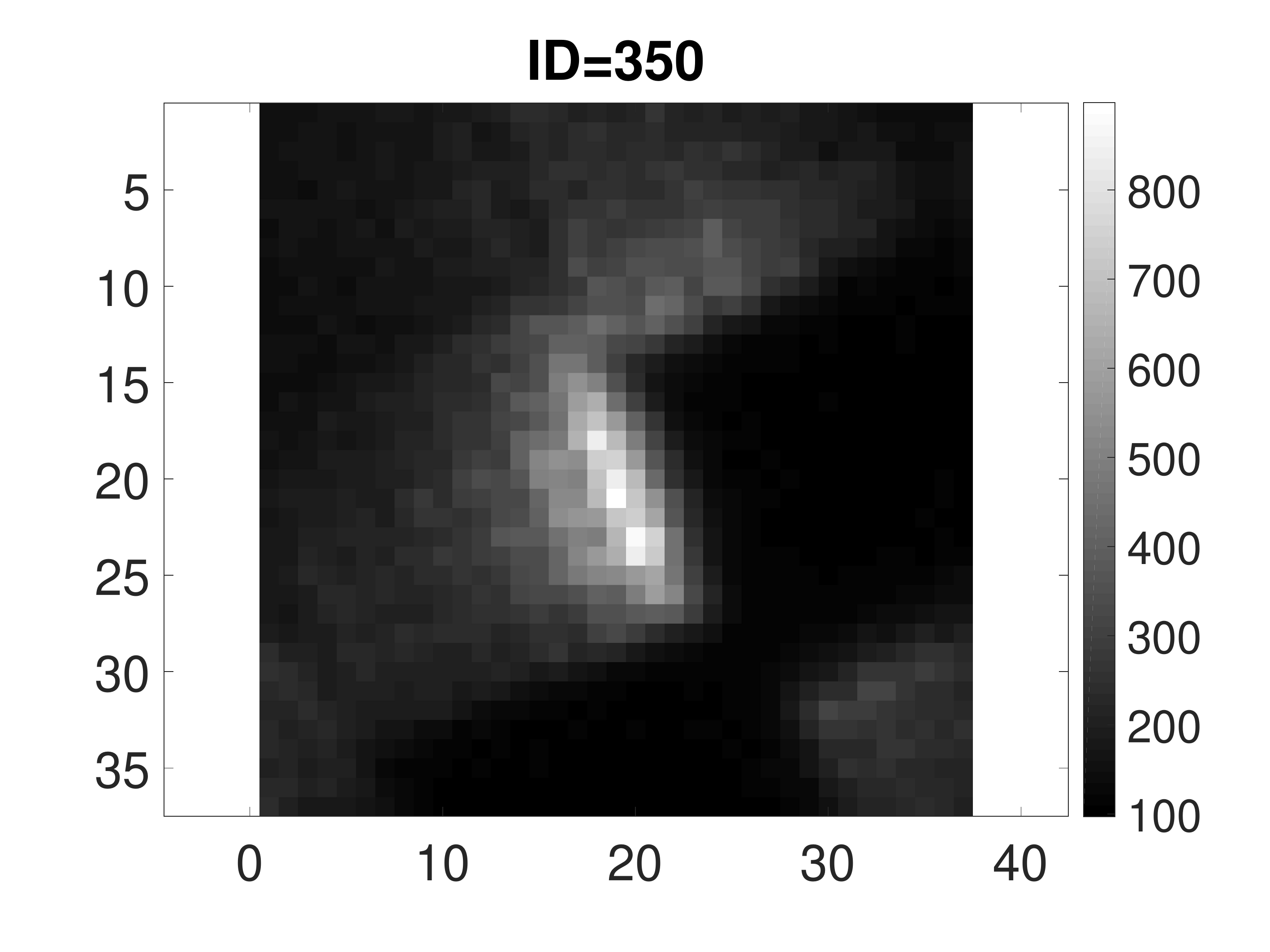}
		\includegraphics[width= 0.45\columnwidth]{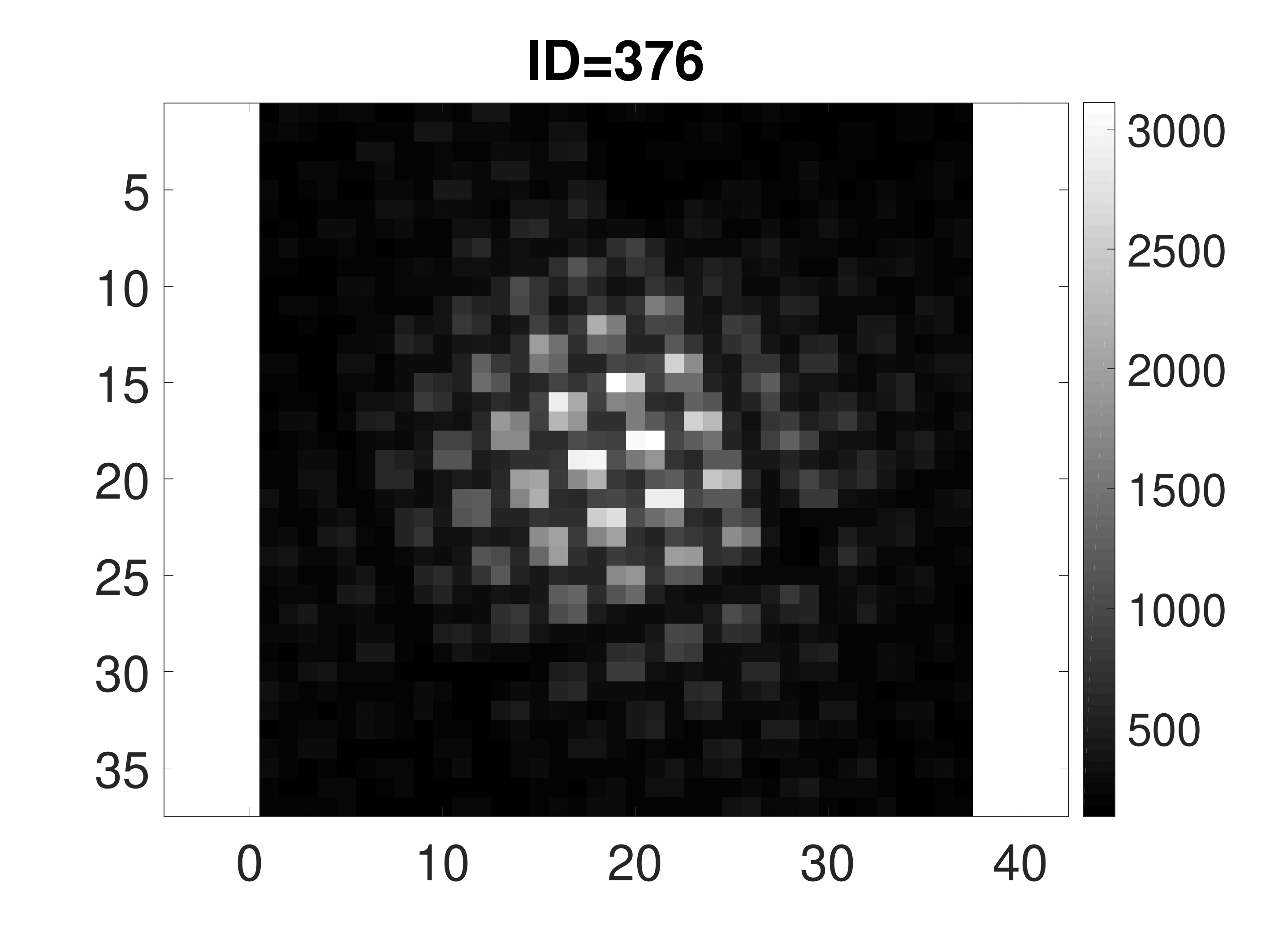}
		\includegraphics[width= 0.45\columnwidth]{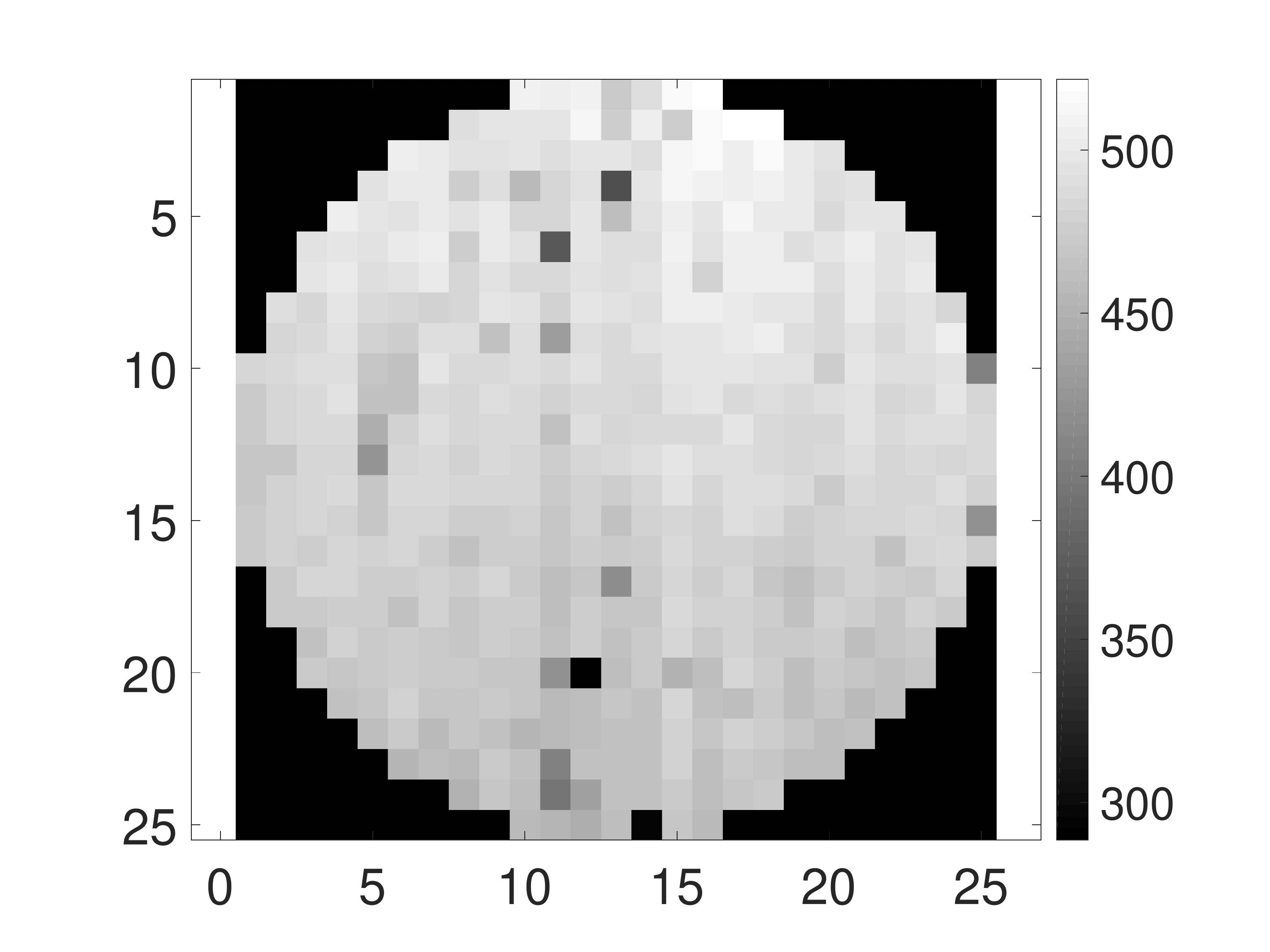}
		\caption{Examples of two poorly sampled spots with low signal counts due to a non-transmitting tile (top); one good sampled spot with high signal counts (bottom-left); Signal-to-noise ratio map for spots of the 25x25 Shack-Hartmann array (bottom-right). The spot signal-to-noise ratio is above 300, typical around 500 for a box size 37x37 pixels. Data taken from Frame\#1 }\label{snr1}
	\end{center}
\end{figure}

Examples of spot sampling and the signal-to-noise ratio for Frame\#1 are shown in Figure \ref{snr1}. We note that from Figure \ref{snr1} that the signal-to-noise ratio is relatively high, typically about 500, so the expected centroid error due to noise factors of the order $\sigma=$0.002 spot diameters. Therefore, from the simulations, we expect the bundle sampling errors to dominate the centroid errors.  From Figure \ref{snr1} we see that the poorly sampled spots are both impeded by a non-transmitting tile that results in a centroid measurement error due to irregular spot shape changes as well as a lower signal-to-noise ratio.

\begin{figure}
	\begin{center}
		\includegraphics[width= 1\columnwidth]{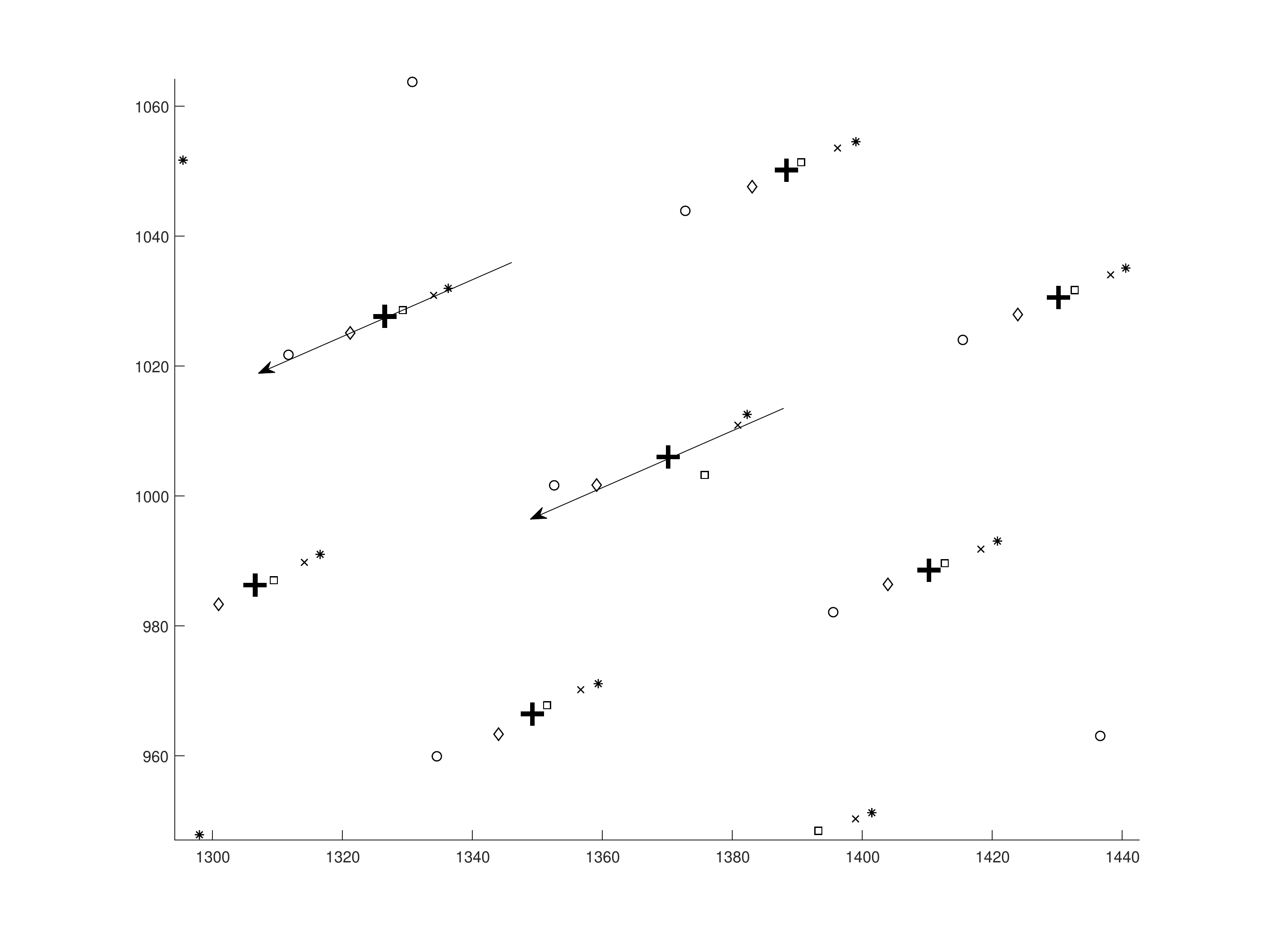}
		\caption{Subset of centroid positions for 5 frames of Shack-Hartmann data. Each frame having a fraction of a spot displacement (approx. North East to South West direction). The displacement vector overlays for a good sampled spot (left) and a poor sampled spot (centre). The plus (+), symbols denote the average centroid or reference frame. The coordinates are in  raw pixels. }\label{SamplingError}
	\end{center}
\end{figure}

To compare with simulations the equivalent Gaussian spot fitted to the microlens theoretical airy disk has a standard deviation of $ \sigma \approx 0.42 \lambda N$, or $ \sigma = $ 12.2$\mu$m. The equivalent spot diameter is approx. 5.81$\sigma$ and encircles 99.6\% of the energy of the Gaussian spot. The diameter of the spot on the detector is  approx. 22 pixels because of the 2x magnification re-imaging lens. The analysis consists of 5 frames each having a unique microlens array displacement as listed in Table \ref{tab:shdisp}. The range in displacement of the spot array pattern being approximately a spot diameter, as shown in Figure \ref{SamplingError}. This was achieved by moving the microlens array in the lateral position using a positioning stage. This allows for the sufficient dithering of the bundle surface in order to quantify the performance in centroid measurement.

\begin{table}
	\caption{Shack-Hartmann relative displacements on bundle.}
	\centering
	\begin{tabular}{@{}c  c  c@{}}
		\hline\hline
		Frame & $\Delta X$ ($\mu$m) &  $\Delta Y$ ($\mu$m) \\ 
		\hline%
		1  & 33.2930 &  15.0079 \\
		2  & 25.2513 &  11.4464 \\ 
		3  & 8.4173  &  3.8198 \\ 
		4  & -18.7698 &  -8.7103 \\ 
		5  & -48.1917 & -21.5638 \\ 
		\hline\hline
	\end{tabular}
	\label{tab:shdisp}
\end{table}

An example of spot centroid measurement errors are shown in Figure \ref{SamplingError}. From Figure \ref{SamplingError} we see that the centroid of the poorly sampled spot for each of the 5 frames deviates from its linear displacement vector unlike its neighbouring sampled spots. This is due a non-transmitting tile that erroneously biases the spot centroid measurement. The smaller deviations of the neighbouring sampled spots in Figure \ref{SamplingError} are due to structure and layout of the wound bundle.

\begin{figure}
	\begin{center}
		\includegraphics[width= 1\columnwidth]{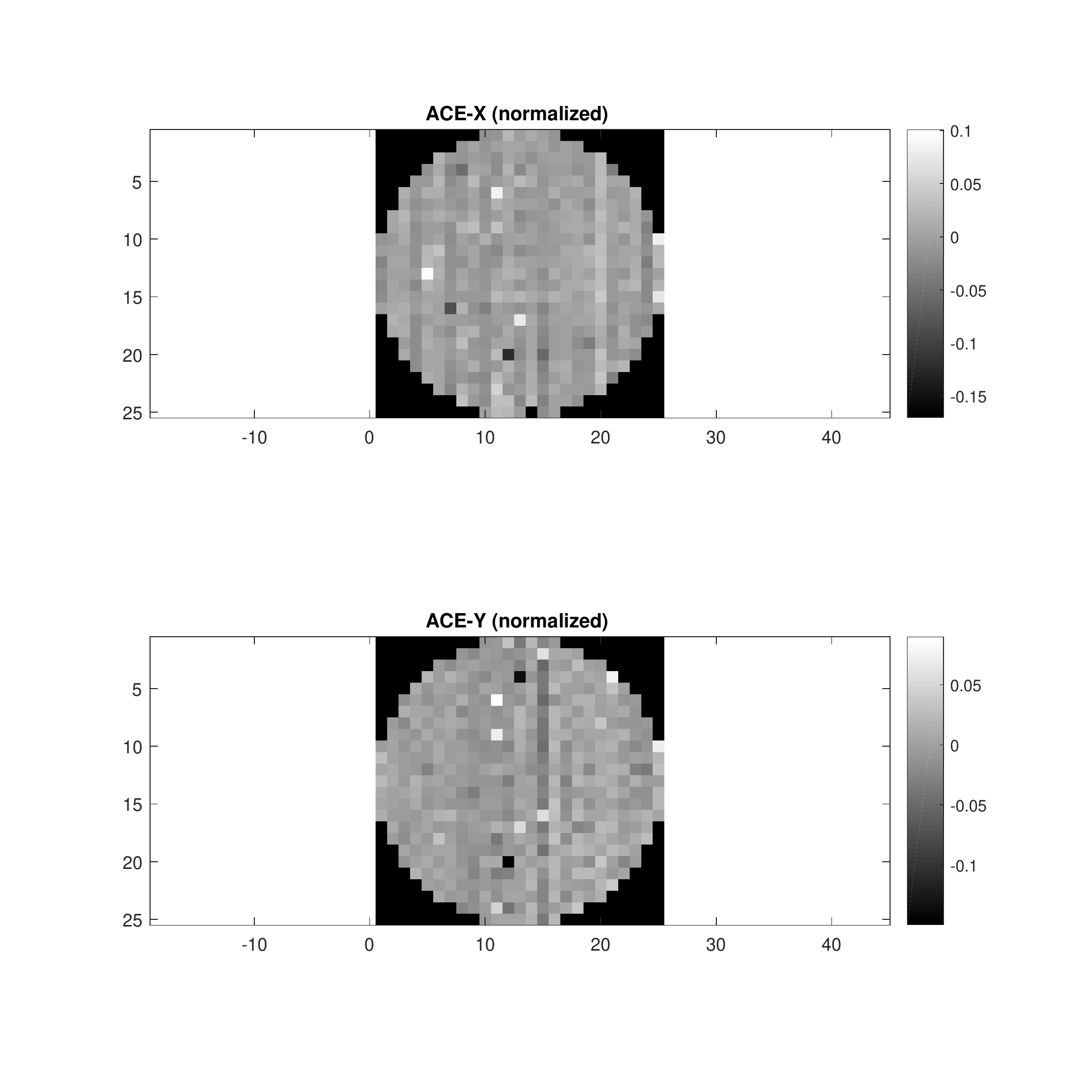}
		\caption{Absolute Centroid Error (ACE) for Frame\#1 in X direction (top) and Y-direction (bottom). Error is normalized by the spot diameter. The X,Y axis units are in lenslet spacings. }\label{SamplingErrorWFS1}
	\end{center}
\end{figure}

\begin{figure}
	\begin{center}
		\includegraphics[width= 1\columnwidth]{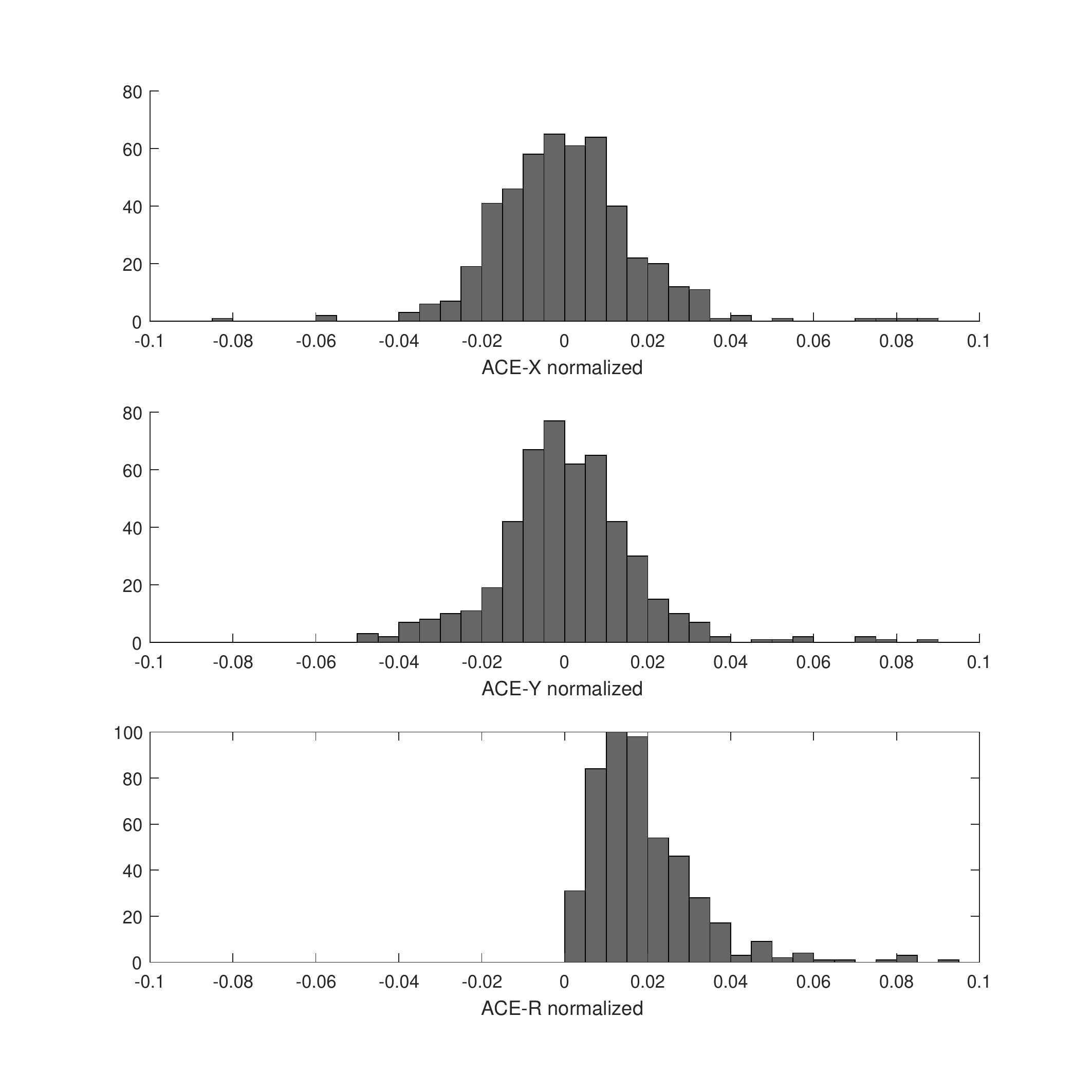}
		\caption{Histogram of Absolute Centroid Error (ACE) for Frame\#1 in X direction (top), Y-direction (middle) and $R=\sqrt{X^2+Y^2}$ (bottom). Error is normalized by the spot diameter. }\label{SamplingErrorWFS1_hist}
	\end{center}
\end{figure}

To see how these sampling issues affect the Shack-Hartmann array, consider the Absolute Centroid Error (ACE) with example shown in Figure \ref{SamplingErrorWFS1}. The ACE is calculated by first subtracting the reference (local spot averages from all frame data) from each individual dataset. It is assumed that reference frame averages out the sampling errors. Next the  global tip-tilt displacement is subtracted by calculating and subtracting the mean displacement of all spots in the array. The residual displacement of individual spot are then normalized by the theoretical spot diameter. The normalization being a useful comparison tool for other systems. A histogram of the ACE computed from Figure \ref{SamplingErrorWFS1} are shown in Figure \ref{SamplingErrorWFS1_hist}. The ACE standard deviation for X and Y ($\sigma_x$ and $\sigma_y$) are listed in Table \ref{tab:sdace}. We get a mean $\sigma_x$=0.0240 and $\sigma_y$=0.0215 that can be compared to the simulation results shown in Figure \ref{Result7} to Figure \ref{Result11}. From Table \ref{tab:sdace} we note a similarity in the values for the ACE standard deviation, with Y being slightly less than X, except for Frame\#5. A reason for the lower Y values could be the relative alignment of the displacement vector to the wound bundle row/col structure. 

\begin{table}
	\caption{Standard deviation of absolute centroid error.}
	\centering
	\begin{tabular}{@{}c  c  c@{}}
		\hline\hline
		Frame & $\sigma_x$ (spot) &  $\sigma_y$ (spot) \\ 
		\hline%
		1  & 0.0200  &  0.0193 \\
		2  & 0.0197  &  0.0165 \\ 
		3  & 0.0222  &  0.0184 \\ 
		4  & 0.0319  &  0.0244 \\ 
		5  & 0.0264  &  0.0290 \\ 
		mean  & 0.0240  &  0.0215 \\ 
		\hline\hline
	\end{tabular}
	\label{tab:sdace}
	
	\medskip
	\tabnote{$^a$Normalized by dividing centroid by spot diameter. }
\end{table}

The parameters for a stable distribution fit to the ACE for X and Y data are listed in Table \ref{tab:statx} and \ref{tab:staty}. A stable distribution is used as it is suitable for modelling heavy tails and skewness. A stable distribution is helpful in generating errors for fast simulations that model the wound bundle for wavefront sensing applications. The first parameter $\alpha$, is the shape parameter describing the tails. The second parameter $\beta$, is the shape parameter describing the skewness. The third and fourth parameters, $\gamma$ and $\delta$ are used for the scale and location. From Table \ref{tab:statx} and \ref{tab:staty} we note the similarity of $\alpha$ for X and Y being 1.7225 and 1.6535. This gives a distribution with a larger tail (narrower peak) compared to Gaussian distribution which has $\alpha$=2.          

\begin{table}
	\caption{Stable probability distribution parameter for absolute centroid error in $x$.}
	\centering
	\begin{tabular}{@{}c  c c c  c@{}}
		\hline\hline
		Frame & $\alpha$ &  $\beta$ & $\gamma$ & $\delta$ \\ 
		\hline%
		1  & 1.7445  &  0.3366  &  0.0101 &  -0.0009  \\
		2  & 1.6217  &  0.6226  &  0.0090 &  -0.0030  \\
		3  & 1.8834  &  0.2933  &  0.0117 &  -0.0014 \\ 
		4  & 1.7064  & -0.6796  &  0.0151 &   0.0041 \\
		5  & 1.6565  & -0.4803  &  0.0127 &   0.0027 \\
		mean  & 1.7225 &   0.0185 &   0.0117  &   0.0003 \\
		\hline\hline
	\end{tabular}
	\label{tab:statx}
	
	\medskip
	\tabnote{$^a$Normalized by dividing centroid by spot diameter. }
\end{table}

\begin{table}
	\caption{Stable probability distribution parameter for absolute centroid error in $y$.}
	\centering
	\begin{tabular}{@{}c  c c c  c@{}}
		\hline\hline
		Frame & $\alpha$ &  $\beta$ & $\gamma$ & $\delta$ \\ 
		\hline%
		1  & 1.6394  & -0.0549  &  0.0094   &   0.0003 \\ 
		2  & 1.5982  & -0.2850  &  -0.2850  &   0.0082 \\ 
		3  & 1.6908  & -0.4276  &  0.0089   &   0.0020 \\
		4  & 1.7559  &  0.2048  &  0.0125   &  -0.0003 \\ 
		5  & 1.5831  &  0.6168  & 0.0139    & -0.0042 \\
		mean  & 1.6535 &   0.0108  &  -0.0481  &   0.0012 \\
		\hline\hline
	\end{tabular}
	\label{tab:staty}
	
	\medskip
	\tabnote{$^a$Normalized by dividing centroid by spot diameter. }
\end{table}

The wavefront is reconstructed from the Shack-Hartmann centroid data using the modal method using Zernike polynomials. The dimension of the Shack-Hartmann array being a circular pupil with 25 lenslets across the diameter. The number of Zernike modes being limited to 60 modes. The reference for the Shack-Hartmann array being the average of the corresponding local centroid over all frames, as shown in Figure \ref{SamplingError}.

\begin{figure}
	\begin{center}
		\includegraphics[width= 1\columnwidth]{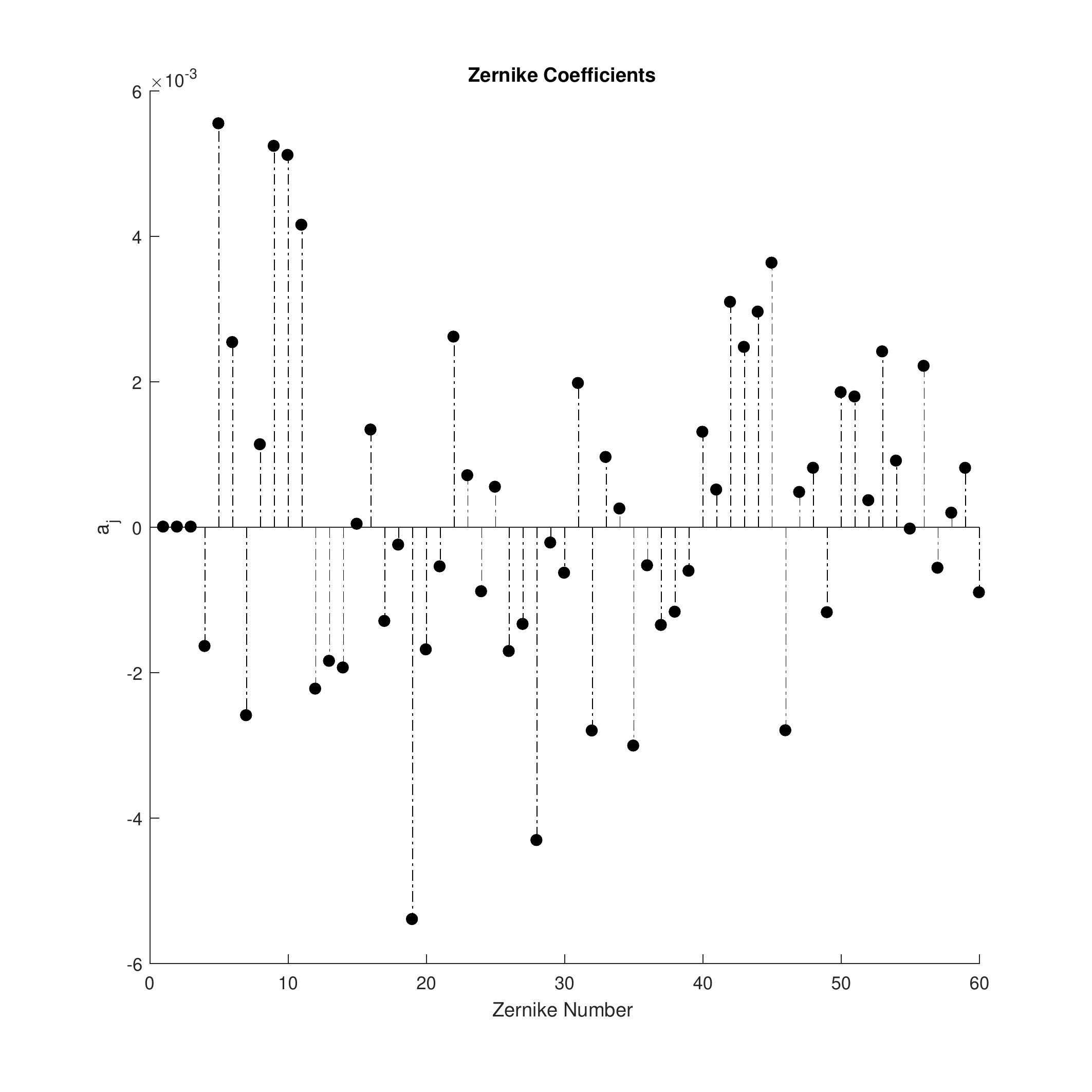}
		\caption{Zernike coefficients for Frame\#1. The tip-tilt coefficients being set to zero. }\label{zernike1}
	\end{center}
\end{figure}

\begin{figure}
	\begin{center}
		\includegraphics[width= 1\columnwidth]{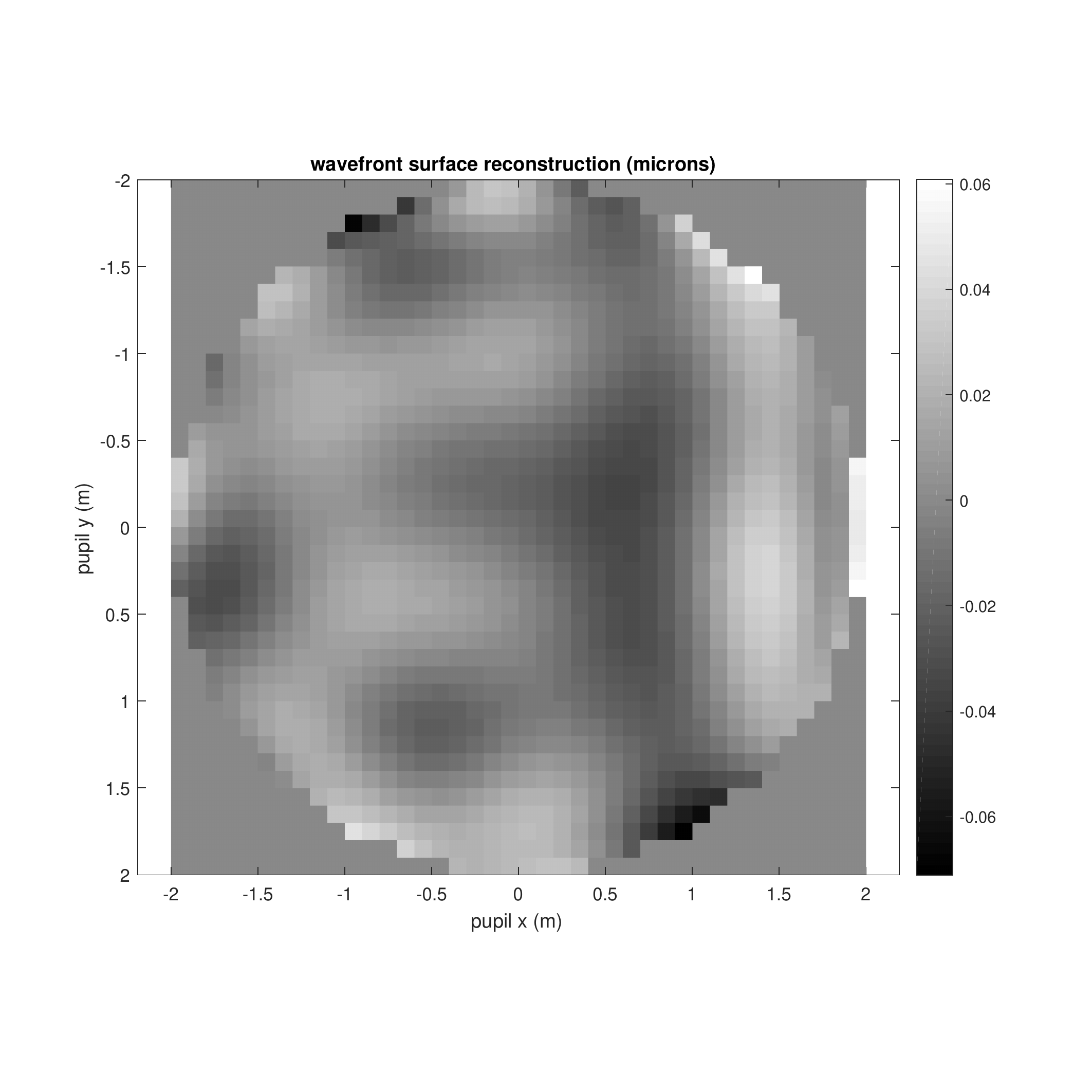}
		\caption{Wavefront (tip-tilt subtracted) for Frame\#1. The RMS is 0.0182 microns. }\label{wavefront1}
	\end{center}
\end{figure}

For each frame the reference centroid data was subtracted and then the wavefront reconstructed. The piston, tip-tilt Zernike terms set to zero (Z1=0, Z2=0, Z3=0) and then the RMS of the wavefront computed. Removing the wavefront tip-tilt component compensates for the displacement of the Shack-Hartmann array allowing the sampling effects to dominate the measured wavefront. The results for the  Frame\#1 is shown in Figure \ref{zernike1} and Figure \ref{wavefront1}. The summary wavefront RMS results can be found in Table \ref{tab:waverr}. The mean RMS of the tip-tilt subtracted wavefront from Table \ref{tab:waverr} is $\sigma$=0.0219 $\mu$m or $\sigma \approx \lambda/30$. We note that the mean RMS is skewed by Frame\#5 having $\sigma$=0.0354 $\mu$m. For a classical Shack-Hartmann array with a CCD we expect the RMS of the tip-tilt subtracted wavefront for different displacements to be near zero (i.e. no sampling effects).

\begin{table}
	\caption{Shack-Hartmann reconstructed wavefront.}
	\centering
	\begin{tabular}{@{}c  c@{}}
		\hline\hline
		Frame & RMS ($\mu$m) \\ 
		\hline%
		1  & 0.0182\\
		2  & 0.0149\\
		3  & 0.0200\\
		4  & 0.0210\\
		5  & 0.0354\\
		mean & 0.0219 \\
		\hline\hline
	\end{tabular}
	\label{tab:waverr}
	
	\medskip
	\tabnote{$^a$Reference wavefront is mean of datasets. }
	\tabnote{$^b$Global tip-tilt terms subtracted from each dataset. }
	
\end{table}

\section{CONCLUSIONS}

In this paper, we have investigated the feasibility of using a wound fibre image bundle to relay the image formed by a Shack-Hartmann wavefront sensor. The key advantage being that the wound bundle  facilitates a compact wavefront sensor design that can fit into a Starbug built for the upcoming extremely large telescopes. The application of the technology allows the efficient positioning of many wavefront sensors over the focal plane for MOAO. We are particularly interested in seeing how this technology can benefit the Giant Magellan Telescope.

We have provided a description of the experimental setup and the limitations expected from the wound fibre image bundle technology, such as the 'chicken wire' and spot blemishes (non-transmitting fibres). We have characterized the performance of a demo wound bundle, provided by Schott North America.

The throughput of the demo wound fibre image bundle measured from 40\% to 50\% over 400 nm to 1600 nm. This is expected decrease slightly with increased length but provides the flexibility for cable management (i.e. 4 m lengths should be practical). The transmission in the near-infrared and improved performance of low-noise high-read out detectors allows the prospect for near-infrared wavefront sensing (i.e. wavelengths around 1600 nm should be practical).

Simulations have been performed to understand the performances of the demo wound fibre for wavefront sensing. The simulations clearly show that for a low signal-to-noise ratio (i.e. less than 30) the wound fibre imaging bundle is not the limiting factor. It is only at high signal-to-noise ratios that the 'chicken wire' limitation is apparent and sets the overall centroid error noise floor. It is also shown that the spot blemish plays an important role for the centroiding measurement accuracy. It can increase the SD of ACE significantly. It is necessary to define the tolerance of the spot blemish within the wound bundle. Our simulation shows that the tolerance less than 1$\%$ will be needed. While the tolerance for chicken wire width is less tight. 

Experiments performed on the demo wound fibre image bundle examined the (i) non-transmitting tiles (spot blemishes); (ii) misaligned tiles (distortion) and the (ii) wavefront measurement. This allows us to examine that which are not easily modelled through simulations. We measured approx. 35 full and 12 partial non-transmitting tiles, which total less than 1\% of the imaging surface. Misaligned tiles between front and back imaging surfaces are much fewer in number, the largest of misalignment being around 10 microns. Keeping in mind that the commercial bundle has much better specifications (less than 6 blemishes for an equivalent sized bundle).

We measured the normalised centroid error of the demo wound fibre image bundle  to have a mean standard deviation, $\sigma_x$=0.0240 and $\sigma_y$=0.0215 of the spot diameter. These values are similar to the simulation for the Y component. We have reported the parameters for stable distribution fits to the normalised centroid errors. The stable distributions can be used to model the centroid errors to speed up simulations. 

The mean RMS of the resulting wavefront is measured to be $\sigma$=0.0219 $\mu$m or equivalently $\sigma \approx \lambda/30$. Subject to further investigation, it may be possible with calibration to further reduce the wavefront error induced by the wound bundle. However, this level of error should be acceptable for most wavefront sensing applications.

We therefore conclude that the use of a wound fibre imaging bundle is feasible for Shack-Hartmann wavefront sensing. Further, incorporating the wavefront sensor into a Starbug for focal plane positioning provides enhanced science opportunities for future astronomical instrumentation. We are now planning the development of a Starbug wavefront sensor suitable for large telescopes.

\bibliographystyle{pasa-mnras}
\bibliography{AccuracyShackHartmannWavefrontSensorCoherentFibeeImageBundle}

\end{document}